\documentclass[aps,prx,preprint,onecolumn,citeautoscript,nofootinbib,eqsecnum,superscriptaddress]{revtex4-2}
\usepackage[T1]{fontenc}
\usepackage{feynmp}
\usepackage{comment}
\usepackage{subcaption}
\usepackage{amsmath}
\usepackage{amssymb}
\usepackage{amsthm}
\usepackage{amsfonts}
\usepackage{bm,bbm}
\usepackage{braket}
\usepackage{xcolor}
\usepackage{pifont}
\usepackage{slashed}
\usepackage[mathscr]{euscript}
\usepackage[shortlabels]{enumitem}
\usepackage{tikz-feynman}
\usepackage[papersize={8.5in,11in}]{geometry}
\usepackage[colorlinks=true]{hyperref}
\hypersetup{
    bookmarks=true,         
    unicode=false,          
    pdftoolbar=true,        
    pdfmenubar=true,        
    pdffitwindow=false,     
    pdfstartview={FitH},    
    pdfsubject={},   
    pdfcreator={},   
    pdfproducer={}, 
    pdfkeywords={} {} {}, 
    pdfnewwindow=true,      
    colorlinks=true,       
    linkcolor=magenta, 
    citecolor=blue,        
    filecolor=magenta,      
    urlcolor=blue           
} 

\usepackage{amsfonts}
\usepackage{upgreek}
\usepackage{slashed}
\usepackage{latexsym}

\newcommand{\beq}{\begin{eqnarray}}
\newcommand{\eeq}{\end{eqnarray}}

\usepackage{mathtools}

\newcommand{\bsp}{\begin{split}}
\newcommand{\esp}{\end{split}}

\newcommand{\sgn}{{\rm sgn}}

\newcommand{\be}{\begin{equation}}
\newcommand{\ee}{\end{equation}}

\graphicspath{{./figures/}}
\def\bea{\begin{eqnarray}}
\def\eea{\end{eqnarray}}

\begin{document}
\preprint{\href{https://arxiv.org/abs/2305.02336}{arXiv:2305.02336}}

\title{Theory of shot noise in strange metals}
\author{Alexander Nikolaenko}
\affiliation{Department of Physics, Harvard University, Cambridge, MA 02138, USA
}

\author{Subir Sachdev}
\affiliation{Department of Physics, Harvard University, Cambridge, MA 02138, USA
}

\author{Aavishkar A. Patel}
\affiliation{Center for Computational Quantum Physics, Flatiron Institute, New York,
New York, 10010, USA}

\date{\today
\\
\vspace{0.4in}}

\begin{abstract}
We extend the theory of shot noise in coherent metals to shot noise in strange metals without quasiparticle excitations. This requires a generalization of the Boltzmann equation with a noise source to distribution functions which depend independently on the excitation momentum and energy. We apply this theory to a model of a strange metal with linear in temperature ($T$) resistivity, describing a Fermi surface with a spatially random Yukawa coupling to a critical boson. We find a suppression of the Fano factor in the strange metal, and describe the dependence of the shot noise on temperature and applied voltage. At low temperatures, we obtain a Fano factor equal to $1/6$, in contrast to the $1/3$ Fano factor in diffusive metals with quasiparticles. Our results are in general agreement with recent observations by Chen {\it et al.} (arXiv:2206.00673). We further compare the random Yukawa model to quasi-elastic electron-phonon scattering that also generates $T$-linear resistivity, and argue that shot noise observations offer a useful diagnostic to distinguish between them.
\end{abstract}

\pacs{Valid PACS appear here}
\maketitle{}
\newpage

\section{Introduction}
\label{sec:intro}

Strange metals have long intrigued solid state physicists, due to their incompatibility with the standard Landau Fermi liquid theory of metals, and their apparent lack of well-defined quasiparticle excitations that carry electrical charge \cite{SachdevRMP22}. This, most famously, manifests in a linear-in-temperature ($T$) resistivity as $T\rightarrow0$ as opposed to the classic $T^2$ dependence of the resistivity in ordinary Fermi liquid metals, along with large electron-electron scattering rates inferred from transport measurements that approach a `universal' value of $k_B T/\hbar$ \cite{Hartnoll:2021ydi}. The near-ubiquitous observation of strange metal resistivity at low temperatures across several different materials with strongly correlated electrons, such as cuprate \cite{Legros2019} and pnictide \cite{Hayes2016} superconductors, heavy fermion materials \cite{Nguyen2021}, iron selenide \cite{Jiang2023}, and twisted bilayer graphene \cite{Jaoui2022} suggests a universal underlying mechanism that is not too sensitive to material details. Unfortunately however, apart from resistivity measurements, there are only a few other experimental probes such as optical conductivity \cite{Michon2022} and photoemission spectroscopy \cite{Nakajima2020} that illuminate the nature of electron-electron scattering processes in strange metals, leaving much left to be done in terms of constructing a complete picture of them from experimental observations alone.   

Amidst this drought, a recent experimental breakthrough \cite{Natelson23} has shed new light on strange metal phases, by measuring shot noise in electrical conduction. Shot noise is a fundamental consequence of the granularity of a stream; if any quantity is transported in discrete packets (such as electrical charge being transported by discrete electrons or quasiparticles), uncertainties due to the Poisson distribution, which describes the occurrence of independent random events, lead to significant fluctuations in the output when the current is small. For non-interacting electrons moving freely through space, the spectral power density $S$ of the noise is given by $S=2eI$, where $e$ is the electron charge and $I$ is the average current. When the motion of the electrons is impeded by impurities but the electrons are still non-interacting, $S$ is suppressed to $1/3$ of this value \cite{Beenakker92, Nagaev92, deJongthesis, deJongBeen}. A dimensionless ratio called the Fano factor $F=S/(2eI)$ \cite{Kobayashi:2021gii} is often used to characterize the intensity of shot noise in electrical transport measurements. The experiment of Ref. \cite{Natelson23} showed that the shot noise in a heavy fermion strange metal wire is suppressed by a Fano factor which is significantly smaller than the expected value of $1/3$ in a non-interacting electron gas with impurity scattering, and even smaller than the expected value of $\sqrt{3}/4$ in the strongly interacting regime \cite{Kozub95,Nagaev95,SiNoise}. 

In a strange metal, electrical charge is not discretized into well defined electron quasiparticles, but is instead smeared across a dense continuum of highly entangled low energy quantum many-body states \cite{SachdevRMP22}. This occurs because of strong inelastic electron-electron scattering, that blends quasiparticles of different energies together into an entangled many-body soup. In contrast, elastic scattering such as electron-impurity scattering conserves quasiparticle energies and only rotates the microcanonical basis of single-quasiparticle excitations; the factorizability of the many-body states into individual single-particle states is thus preserved. Consequently, due to the reduced or nonexistent granularity of the charge stream in a a strange metal, one would expect the Fano factor to be strongly suppressed in a strange metal relative to a metal with well-defined quasiparticles. 

This paper will develop a theoretical framework for computing shot noise in a metal without quasiparticle excitations. The earlier computations of shot noise in the presence of quasiparticles extended the conventional Boltzmann equation to a `noisy' Boltzmann equation which had a noise source term along with a collision term \cite{Kogan69,Green06}. This noisy Boltzmann equation describes the fluctuations in the particle distribution function $\delta f$, which is a function of the momentum of the quasiparticles $\mathbf{k}$, along with a dependence upon a spatial position $\mathbf{r}$ (we will assume steady-state situations, and so there is no dependence on time, $t$). But in a metal without quasiparticles, excitations with momentum $\mathbf{k}$ do not have a definite energy as a consequence of the inelastic scattering mentioned in the previous paragraph. Such metals have therefore been analyzed by a Boltzmann equation in which the distribution function has a separate dependence upon $\mathbf{k}$ and the excitation energy $\omega$ \cite{PrangeKadanoff,NaveLee07,Patel:2017mjv}, but in the absence of noise sources. Our theory of shot noise in strange metals will combine these two sets of earlier results to obtain a noisy Boltzmann equation in which $\delta f$ depends upon both $\mathbf{k}$ and $\omega$.

We will apply such a noisy Boltzmann equation to the recently developed universal theory of strange metals \cite{Aldape20,Patel21,Guo2022,Patel:2022gdh}, which consists of fermions at finite density with a spatially disordered Yukawa coupling to quantum critical scalars, the latter representing order parameters or particles with emergent gauge charges, in two spatial dimensions. This theory, which is exactly solvable in a large $N$ limit and is expected to be valid even beyond it \cite{Patel21}, has several appealing features in the context of experiments on strange metals:
\begin{itemize}
    \item {\it Simplicity}: the theory of fermions at finite density coupled to fluctuating scalars \cite{Hertz1976, Millis1993} is the natural extension of standard mean field theories of ordered phases in which the scalars are condensed. The universal theory of strange metals is simply a careful extension of this picture to disordered systems (all materials that show strange metal behavior have some amount of disorder).
    \item {\it Non-trivial DC resistivity}: as is well known \cite{holographicQM}, at finite charge density, the current relaxation that is required to obtain a nonzero DC resistivity needs momentum relaxation. In the limit $T\rightarrow0$, sufficient momentum relaxation cannot be provided by either phonons or Umklapp processes, leaving disorder as the only general and realistic option to obtain strange metal behavior. The universal theory of strange metals accounts for the spatial modulation of electron-electron interactions by disorder, a naturally expected feature of strongly correlated systems that was nevertheless overlooked in earlier work. This spatially modulated interaction leads to disorder-induced {\it inelastic} electron scattering, and therefore temperature dependence of the resistivity. This is to be contrasted from the usual temperature-independent `residual' scattering rate that arises from {\it elastic} electron-impurity scattering. 
    \item {\it Transport and thermodynamic properties:} The universal theory of strange metals in two spatial dimensions leads to scattering rates that are linear in temperature $T$ and frequency $\omega$ (along with $\omega/T$ scaling) in {\it both} single particle properties and transport, as well as a specific heat that scales as $T\ln T$, thereby reproducing the `marginal Fermi liquid' \cite{MFL1989} phenomenology of strange metals seen in experiments such as Refs. \cite{Legros2019, Nakajima2020, Michon2022, Girod2021}. In the limit of strong spatially disordered coupling, the theory also produces transport scattering rates that are $\mathcal{O}(k_B T/\hbar)$ when defined with respect to the renormalized fermion mass. This phenomenon is also experimentally observed and is known as `Planckian dissipation' \cite{Hartnoll:2021ydi}. 
    \item {\it Universality:} The theory produces the above hallmarks of strange metal behavior irrespective of the type of order parameter that the critical scalar represents, as long as the system is two dimensional or quasi-two dimensional with respect to the critical scalar fluctuations. The scalar can also be an emergent `slave boson' inducing a Fermi volume changing transition \cite{Analytis22,Aldape20}, as we review here in Appendix~\ref{app:KLM}. It can therefore explain why strange metal behavior is seen at finite temperature above quantum critical points in a variety of materials \cite{Legros2019, Hayes2016, Nakajima2020, Nguyen2021, Jiang2023, Steglich00} that are quasi-two dimensional, but have different kinds of quantum phase transitions.
\end{itemize}

In this universal theory of strange metals, the Yukawa coupling is assumed to have the form $g + g'(\mathbf{r})$, where $g$ is spatially independent, while $g'(\mathbf{r})$ is spatially random with zero mean and variance $g'^2$. A random potential $v(\mathbf{r})$, with zero mean and variance $v^2$ is also applied to the fermions, as is usual in the theory of disordered metals. It was shown that the elastic scattering from the random potential determined a zero temperature residual resistivity proportional to $v^2$, the singular inelastic scattering from the coupling $g$ cancelled out in the resistivity \cite{Hartnoll:2007ih,Hartnoll:2014gba,Maslov17a,Maslov17b,Guo2022,Patel:2022gdh,Shi22}, and only the inelastic scattering 
from the spatially random coupling $g'$ led to strange metal behavior in the conductivity \cite{Patel:2022gdh}, including the linear-in-$T$ resitivity. We expect a similar sub-dominance of the singular effects of $g$ in the shot noise (see Sec. \ref{sec:discussion} for details), and so we will only consider here the noisy Boltzmann equation in the presence of $g'$ and $v$. The absence of a $g$ coupling also helps reduce the complexity of the rather involved analysis of the noisy Boltzmann equation dependent upon both $\mathbf{k}$ and $\omega$. As we will see below, the $g'$-$v$ theory is in good accord with the recent shot noise experiment \cite{Natelson23}, producing a Fano factor that is significantly smaller than $1/3$.

A notable feature of our theory is that the boson distribution function is unaffected by the applied electric field. This is a consequence of the dependence of the boson self energy only upon the {\it local\/} fermion Green's functions, and the linear influence of the electric field vanishes upon an angular average. An important consequence is that the electron and boson distribution functions do not thermalize at a common space-dependent temperature and lead to the associated shot noise \cite{Kozub95,Nagaev95,SiNoise}.

The shot noise in our theory is caused by a Poisson process, whose variance ({\it i.e.} noise power) is proportional to its mean ({\it i.e.} current), regardless of whether the system is a non-Fermi liquid or not. Indeed, the Poissonian property of the proportionality of the variance to the mean is preserved by Eqs. (\ref{eq:noiseinout}--\ref{eq:noisemeans1}) describing the current fluctuations, despite the additional dependence on frequency $\omega$ in the non-Fermi liquid. Therefore, the noise should remain proportional to the average current regardless of whether the system is a non-Fermi liquid or not. Furthermore, because the boson remains in equilibrium as discussed above, additional voltage and current dependencies into the scattering processes and therefore the noise are not introduced. 

The rest of this paper is organized as follows: in Sec. \ref{sec:overview}, we provide the theoretical model and summarize the main results of the shot noise computation using it. In Sec. \ref{sec:mainstuff} we describe the generalization of the noisy Boltzmann equation formalism to metals without quasiparticles, and perform the computation of shot noise in the universal theory of strange metals. We conclude in Sec. \ref{sec:discussion}, giving physical context to our model and computations. We also contrast the universal theory of strange metals with quasi-elastic electron-phonon scattering, both of which produce $T$-linear resistivity (although not as $T\rightarrow0$ for the latter), but, as we will show, have very different shot noise signatures that can be used to differentiate between them.

\section{Overview}
\label{sec:overview}

\subsection{Model of strange metals}

We will study the $g'$-$v$ model of fermions $c$ and bosons $\phi$ defined by the imaginary time ($\tau$) Lagrangian $\mathcal{L}_c  + \mathcal{L}_\phi + \mathcal{L}_v + \mathcal{L}_{g'}$
\begin{align}
\mathcal{L}_c & = \sum_{\mathbf{k}} c_{\mathbf{k}}^\dagger (\tau) \left( \frac{\partial}{\partial \tau} + \epsilon_\mathbf{k} \right) c_\mathbf{k} (\tau) \nonumber \\
\mathcal{L}_\phi & = \frac{1}{2} \int d^2 \mathbf{r} \, \left[ \left( \partial_\tau \phi (\mathbf{r}, \tau) \right)^2 + \left( \mathbf{\nabla}_\mathbf{r} \phi (\mathbf{r}, \tau) \right)^2  + m_b^2 \left(\phi (\mathbf{r}, \tau) \right)^2 + \ldots \right] \nonumber \\
\mathcal{L}_{v} & =  \int d^2 \mathbf{r} \,  v(\mathbf{r}) c^\dagger (\mathbf{r},\tau) c  (\mathbf{r},\tau)  \nonumber \\
\mathcal{L}_{g'} & =  \int d^2 \mathbf{r} \, g'(\mathbf{r}) \, c^{\dagger}(\mathbf{r},\tau)  c(\mathbf{r},\tau) \, \phi(\mathbf{r},\tau)\,.\label{f4}
\end{align}
The spatially random couplings $v(\mathbf{r})$ and $g'(\mathbf{r})$ are non-dynamical and so independent of $\tau$; they obey the spatial averages
\begin{align}
\overline{v (\mathbf{r})} & = 0 \quad, \quad \overline{v(\mathbf{r}) v(\mathbf{r'})} = v^2 \delta(\mathbf{r}-\mathbf{r'}) \nonumber \\
\overline{g'(\mathbf{r})} & = 0 \quad, \quad \overline{g'(\mathbf{r}) g'(\mathbf{r'})} = g'^2 \delta(\mathbf{r}-\mathbf{r'}). \label{f7}
\end{align}
Such a model has been studied in earlier work \cite{Aldape20,Patel21,Guo2022,Patel:2022gdh} using a Sachdev-Ye-Kitaev (SYK)-like large $N$ expansion obtained by endowing the fields and couplings with a random dependence on an additional flavor index: this led to a theory whose large $N$ saddle-point was equivalent to solving the fully self-consistent Migdal-Eliashberg equations. Here, we shall directly examine the model without flavor indices, and obtain equivalent results by applying the canonical Kadanoff-Baym analysis \cite{kadanoff1962} to the Migdal-Eliashberg self energies.

As in previous work \cite{Patel21}, we also need to account for the boson self-interactions near the quantum critical point, which are not explicitly written out in $\mathcal{L}_\phi$. The bosons remain in thermal equilibrium in the presence of the applied voltage, and then the retarded local boson Green's function is given by
\begin{equation}
D^R_\mathrm{loc}(\Omega) = \int_0^{\Lambda_q}\frac{q dq}{2\pi}\frac{1}{q^2-ic_d\Omega+m^2(T)},~~c_d = \left(\frac{k_F}{v_F}\right)^2 \frac{{g'}^2}{8 \pi}\,,
\label{eq:boson_prop}
\end{equation}
where $v_F$ is the Fermi velocity, $\Lambda_q$ is a UV momentum cutoff, and $k_F$ is the Fermi wavevector (we are assuming a circular Fermi surface and a quadratic fermion dispersion). As a consequence of the self-interactions, the renormalized boson `mass' $m^2(T)$ is determined by solving (on the Matsubara axis)
\begin{equation}
T\sum_{\Omega_m}\int_0^{\Lambda_q}\frac{q dq}{2\pi}\frac{1}{q^2+c_d|\Omega_m|+m^2(T)} - \int\frac{d\Omega}{2\pi}\int_0^{\Lambda_q}\frac{q dq}{2\pi}\frac{1}{q^2+c_d|\Omega|} = \Delta\kappa,
\label{eq:thermal_mass}
\end{equation}
where $\Delta\kappa = 0$ at the quantum critical point (QCP), which we will consider in most of the rest of this paper. At the QCP $m^2(T)\sim c_d T$ up to logarithmic corrections \cite{Patel:2014jfa}. We can also tune away from the QCP by setting $\Delta\kappa <0$; in this case $m^2(T)$ doesn't vanish as $T\rightarrow 0$, and the resulting `soft gapped' boson is a weaker scatterer of electrons, leading to a $T^2$ resistivity instead of a $T$-linear resistivity at low $T$ \cite{Aldape20}. Shot noise for $\Delta \kappa < 0$ is described in Appendix~\ref{app:NC}.

The local boson spectral function, 
\beq
B_\mathrm{loc}(\Omega)=2\mathrm{Im}[D^R_\mathrm{loc}(\Omega)],
\label{eqn:bloc}
\eeq
will play an important role in our analysis.

\subsection{Main results}

Given the lengthy analysis to follow, it is useful to summarize our main results. The shot noise is measured by the noise power $S$, defined by 
\begin{equation}
S = \int dt\int d^2\mathbf{r}~\langle \delta \mathbf{j}(\mathbf{r},t)\cdot \delta\mathbf{j}(0,0)\rangle.
\label{eq:noise}
\end{equation}
where $\delta \mathbf{j}(\mathbf{r},t)$ is the fluctuating current in the presence of an applied voltage. For a wire of length $L$, width $W$, at a temperature $T$, and with an applied voltage $V$, we obtain
\begin{equation}
\begin{split}
 &S=-\frac{2 \sigma_r W T}{L}  \int d\omega \frac{\partial f_L/\partial \omega+\partial f_R/\partial \omega}{(1- \pi g'^2 s(\omega)/v_F)^2}+\frac{2 \sigma_r W}{3L}  \int d \omega \frac{(f_L(\omega)-f_R(\omega))^2}{(1-\pi g'^2 s(\omega)/v_F)^2}+\\
 &+\frac{2 g'^2 \sigma_r}{2\pi v^2 L } W \int  d\omega d\omega'  \left[ \frac{ B_\mathrm{loc}(\omega'-\omega)}{(1-\pi g'^2 s(\omega)/v_F)^2}+\frac{ B_\mathrm{loc} (\omega'-\omega)}{(1-\pi g'^2 s(\omega')/v_F)^2} \right]  n_B(\omega'-\omega) \\
 &~\times\left(\frac{f_L (\omega)+f_R (\omega)}{2}-\frac{f_L (\omega) f_L (\omega') +
 f_R (\omega) f_R (\omega')}{3}-\frac{f_L (\omega')f_R (\omega)+f_L (\omega) f_R (\omega')}{6}\right)\,.
 \end{split} \label{eq:mainresult}
\end{equation}
Here the `residual' conductivity at $T=0$ (whose inverse is the residual resistivity) is
\begin{equation}
    \sigma_r = \frac{e^2 v_F^2}{4 \pi v^2}\,,
\end{equation}
$f_L$ and $f_R$ are the Fermi functions at the left and right ends of the wire
\begin{equation}
    f_L (\omega) = \frac{1}{e^{(\omega - eV)/T} + 1} \quad, \quad f_R (\omega) = \frac{1}{e^{\omega/T} + 1}\,, \label{fLR}
\end{equation}
$n_B$ is the Bose function
\begin{equation}
    n_B (\omega) = \frac{1}{e^{\omega/T} - 1}\,,
\end{equation}
and $s(\omega)$ is given by a convolution of the boson spectral function and Fermi and Bose functions
\begin{equation}
 s(\omega)= -\int \frac{2 v_F}{(2\pi)^2 v^2}d\omega' B_{\mathrm{loc}}(\omega'-\omega)\left[n_B(\omega'-\omega)+n_F(\omega') \right] \,.
 \label{eqn:s_omega}
\end{equation}

We note some limiting cases of Eq.~(\ref{eq:mainresult}). In the absence of a coupling to the boson, $g'=0$, we have
\begin{equation}
 S=\frac{4 \sigma_r W T}{L} +\frac{2 \sigma_r W T}{3L}\left( e V\coth\left( \frac{e V}{2 T}\right)-2 T\right)\,, 
 \label{eqn:s_diffusive}
\end{equation}
which is the well-known result for the noninteracting electron gas with impurity scattering (also known as a diffusive metal with quasiparticles) \cite{Beenakker92,Nagaev92,deJongthesis,deJongBeen}. At small $V/T$, the shot noise power in Eq.~(\ref{eqn:s_diffusive}) reduces to the Johnson–Nyquist thermal noise $S_{th}=4 \sigma_r W T/L$, while at large $V/T$ we obtain the Fano factor
\begin{equation}
    F=\frac{S}{2 \sigma_r e V  W/L}=\frac{1}{3}\,.
\end{equation}
    
In the presence of interactions, with $g'\neq 0$, we obtain at $T=0$ the Fano factor
\begin{equation}
 F=\frac{S}{2\sigma_r W V/L}=\frac{1}{3} \left( \frac{1+\pi g'^2 V/(4(2\pi)^2v^2)}{1+\pi g'^2 V/(2(2\pi)^2v^2)} \right)\,.
 \label{eqn:fano_factor}
\end{equation}
This shows the suppression of the Fano factor to $1/6$ in the limit of strong interactions and quasiparticle destruction. In the opposite limit of $V\ll T$, 
we obtain the Johnson–Nyquist thermal noise $S_{th}=4 \sigma(g',T) W T/L$
in (\ref{JNg}), where $\sigma (g',T)$ is the bulk conductivity in the presence interactions given in (\ref{sigmagp}). 
The interpolation between these limiting cases in the presence of interactions will be discussed in Section~\ref{sec:gp}.

\section{Noisy Boltzmann Equation}
\label{sec:mainstuff}

We begin by describing the noisy Boltzmann equation. As noted in Section~\ref{sec:intro}, we follow the approach of Nave and Lee \cite{NaveLee07} to derive an appropriate Boltzmann equation that does not assume the existence of quasiparticles, and that of Kogan and Shul'man \cite{Kogan69} and de Jong and Beenakker \cite{deJongpaper,deJongthesis,deJongBeen} to include the noise source. 

For a non-Fermi liquid without quasiparticles, a generalized distribution function can be defined in terms of the lesser Green's function on the Keldysh contour ($G^<$), provided that the fermion spectral function is peaked sharply near the Fermi surface (FS) ({\it i.e.} the spectral function $A(\mathbf{k},\omega)$ peaks as a function of $k=|\mathbf{k}|-k_F$ at least as fast as it does as a function of $\omega$). The generalized distribution function $f(\hat{\mathbf{k}},\omega,\mathbf{r},t)$ is given by
\begin{equation}
f(\hat{\mathbf{k}},\omega,\mathbf{r},t) = k_F\int\frac{dk}{2\pi}\left[-iG^<(k,\hat{\mathbf{k}},\omega,\mathbf{r},t)\right],
\end{equation}
where, again, we assume the FS is circular.

The Boltzmann equation without any noise terms is then given by
\begin{eqnarray}
&&\left(1-\frac{\partial \mathrm{Re}\Sigma^R}{\partial \omega}\right)\frac{\partial f}{\partial t} + \left(\frac{\partial\epsilon_\mathbf{k}}{\partial t} + \frac{\partial \mathrm{Re}\Sigma^R}{\partial t}\right)\frac{\partial f}{\partial \omega} \nonumber \\
&&-\left(\frac{\partial\epsilon_\mathbf{k}}{\partial\mathbf{r}}+\frac{\partial \mathrm{Re}\Sigma^R}{\partial \mathbf{r}}\right)\cdot\frac{\partial f}{\partial \mathbf{k}} +\left(v_F\hat{\mathbf{k}}+\frac{\partial\mathrm{Re}\Sigma^R}{\partial \mathbf{k}}\right)\cdot\frac{\partial f}{\partial\mathbf{r}} = I_{\mathrm{coll}}.
\label{eq:BE_orig}
\end{eqnarray}
Here, the $t$ and $\mathbf{r}$ dependencies of the dispersion $\epsilon_\mathbf{k}$ are included to account for electromagnetic fields via $\mathbf{k}\rightarrow \mathbf{k}-\mathbf{A}$, $\mathbf{E}=-d\mathbf{A}/dt$, $\mathbf{B} = \nabla\times\mathbf{A}$ \cite{Patel:2017mjv}; $\Sigma^R$ is the fermion self energy. Because we do not consider the effects of magnetic fields, we set $\partial\epsilon_\mathbf{k}/\partial\mathbf{r}=0$. For the $v-g'$ model, the collision integral is given by $I_\mathrm{coll}=I_\mathrm{coll}^v+I_\mathrm{coll}^{g'}$
\begin{eqnarray}
&& I_\mathrm{coll}^v = v^2\int\frac{d\hat{\mathbf{k}}'}{2\pi}\Bigg[\left(1-f(\hat{\mathbf{k}},\omega)\right)f(\hat{\mathbf{k}}',\omega)-\left(1-f(\hat{\mathbf{k}}',\omega)\right)f(\hat{\mathbf{k}},\omega)\Bigg], \nonumber \\
&& I_\mathrm{coll}^{g'} = \frac{{g'}^2}{2}\int\frac{d\Omega}{2\pi}\int\frac{d\hat{\mathbf{k}}'}{2\pi}\int\frac{d\omega'}{2\pi}B_\mathrm{loc}(\Omega)\Bigg[2\pi\delta(\omega'-\omega-\Omega)\Big\{\left(n_B(\Omega)+1\right) \nonumber \\
&&\times \left(1-f(\hat{\mathbf{k}},\omega)\right)  f(\hat{\mathbf{k}}',\omega') - n_B(\Omega)\left(1-f(\hat{\mathbf{k}}',\omega')\right)f(\hat{\mathbf{k}},\omega)\Big\} \nonumber \\
&&+2\pi\delta(\omega'-\omega+\Omega)\Big\{n_B(\Omega)\left(1-f(\hat{\mathbf{k}},\omega)\right)f(\hat{\mathbf{k}}',\omega') \nonumber \\
&&-\left(n_B(\Omega)+1\right)\left(1-f(\hat{\mathbf{k}}',\omega')\right)f(\hat{\mathbf{k}},\omega)\Big\}\Bigg].
\label{eq:Cint}
\end{eqnarray}
Here $n_B(\Omega)=1/\left(e^{\Omega/T}-1\right)$, and $B_\mathrm{loc}(\Omega)=2\mathrm{Im}[D^R_\mathrm{loc}(\Omega)]$ is the local boson spectral function, since electrons scatter off local bosonic fluctuations in the $g'$ model \cite{Patel:2022gdh}. Furthermore, we have suppressed the $\mathbf{r},t$ dependence of $f$ in the collision integral as it is the same for all occurences of $f$ in the expression. 

The equilibrium solution for $f$ is $f=n_F(\omega)$ where $n_F(\omega)=1/\left(e^{\omega/T}+1\right)$. This solution nullifies the collision integral. Upon adding noise terms to the Boltzmann equation, $f$ shifts by $\delta f$. Linearizing in $\delta f$, we obtain
\begin{equation}
\left(1-\frac{\partial \mathrm{Re}\Sigma^R}{\partial \omega}\right)\frac{\partial \delta f}{\partial t} +\frac{\partial \epsilon_\mathbf{k}}{\partial t}\frac{\partial\delta f}{\partial\omega} +v_F\hat{\mathbf{k}}\cdot\frac{\partial \delta f}{\partial\mathbf{r}}  = \delta I_{\mathrm{coll}} + \delta j.
\label{eq:BElin}
\end{equation}
Here, $\delta j$ is the noise source term whose autocorrelation function will be described in the next section. We assumed that $\Sigma^R$ and the boson are not affected by the noise source; in the $v-g'$ model this follows from the fact that $\Sigma^R$ and the boson Green's function depend only upon momentum-integrated fermion Green's functions, the noise-induced fluctuations of which are tiny. This is because the fluctuations of fermion Green's functions due to the noise source are due to fluctuations in the scattering processes, which affect them via the fermion self-energies. Since the momentum-integrated Green's functions are independent of the self-energies when the Fermi energy is large \cite{Patel:2022gdh}, it follows that their noise-induced fluctuations are also negligible. The equilibrium $\Sigma^R$ is also independent of $\mathbf{k},\mathbf{r},t$, which further simplifies the equation for $\delta f$. It has a marginal-Fermi liquid form that is given by \cite{Patel:2022gdh}
\begin{eqnarray}
&&\Sigma^R(\omega, T=0) = -i\left(\frac{k_F}{v_F}\right)^2\frac{v^2}{2} -\frac{{g'}^2k_F}{8\pi^2v_F}\omega\ln\left(\frac{ie\Lambda_q^2}{c_d\omega}\right), \nonumber \\
&&\Sigma^R(\omega, T>0) = -i\left(\frac{k_F}{v_F}\right)^2\frac{v^2}{2} + i\frac{{g'}^2k_F}{8\pi v_F} T \Bigg[\ln \left(\frac{\Lambda_q^2}{m^2(T)}\right)-2 \Bigg(\ln\Gamma\left(\frac{\Lambda_q^2}{2c_d\pi T}-\frac{i\omega}{2\pi T}+\frac{1}{2}\right)- \nonumber \\
&&\ln\Gamma\left(\frac{m^2(T)}{2 c_d\pi T}-\frac{i\omega }{2\pi T}+\frac{1}{2}\right)-\ln\Gamma\left(\frac{\Lambda_q^2}{2c_d\pi T}\right)+\ln\Gamma\left(\frac{m^2(T)}{2 c_d\pi T}\right)\Bigg)\Bigg],
\label{eq:fse}
\end{eqnarray}
but these expressions will not be needed in our analysis below.

Proceeding, the linearized collision integral is given by $\delta I_\mathrm{coll} = \delta I_\mathrm{coll}^v + \delta I_\mathrm{coll}^{g'}$
\begin{eqnarray}
&&\delta I_\mathrm{coll}^v = v^2\int\frac{d\hat{\mathbf{k}}'}{2\pi}\Bigg[\delta f(\hat{\mathbf{k}}',\omega) - \delta f(\hat{\mathbf{k}},\omega)\Bigg], \nonumber \\ 
&&\delta I_\mathrm{coll}^{g'} = \frac{{g'}^2}{2}\int\frac{d\Omega}{2\pi}\int\frac{d\hat{\mathbf{k}}'}{2\pi}\int\frac{d\omega'}{2\pi}B_\mathrm{loc}(\Omega)\Bigg[2\pi\delta(\omega'-\omega-\Omega)\Big\{n_B(\Omega)\delta f(\hat{\mathbf{k}},\omega)\nonumber \\
&&\times\left(f(\hat{\mathbf{k}}',\omega')-1\right)+n_B(\Omega)f(\hat{\mathbf{k}},\omega)\delta f(\hat{\mathbf{k}}',\omega')\nonumber \\
&&+\left(n_B(\Omega)+1\right)\left(1-f(\hat{\mathbf{k}},\omega)\right)\delta f(\hat{\mathbf{k}}',\omega')-(n_B(\Omega)+1)\delta f(\hat{\mathbf{k}},\omega)f(\hat{\mathbf{k}},\omega')\Big\}\Bigg] \nonumber \\
&&-\frac{{g'}^2}{2}\int\frac{d\Omega}{2\pi}\int\frac{d\hat{\mathbf{k}}'}{2\pi}\int\frac{d\omega'}{2\pi}B_\mathrm{loc}(\Omega)\Bigg[\left(\hat{\mathbf{k}},\omega\leftrightarrow \hat{\mathbf{k}'},\omega'\right)\Bigg].
\end{eqnarray}

\subsection{Noise Source}

In order to solve the Boltzmann equation for current noise, we need to first write down the noise source $\delta j$. Following Ref.~\onlinecite{Kogan69}, we relate $\delta j$ to the terms in the collision integral for the generalized distribution function $f$: The extraneous flux of particles in state $(\hat{\mathbf{k}},\omega)$ is 
\begin{equation}
\delta j(\hat{\mathbf{k}},\omega,\mathbf{r},t) = \int\frac{d\hat{\mathbf{k}}'}{2\pi}\int\frac{d\omega'}{2\pi}\left[\delta j(\hat{\mathbf{k}}',\omega',\hat{\mathbf{k}},\omega,\mathbf{r},t)-\delta j(\hat{\mathbf{k}},\omega,\hat{\mathbf{k}}',\omega',\mathbf{r},t)\right].
\label{eq:noiseinout}
\end{equation}
As before we will suppress the $(\mathbf{r},t)$ indices when convenient as they are the same for all occurences. The fluctuations in the transition rates (RHS of Eq. (\ref{eq:noiseinout})) come from a Poisson distribution whose variance is equal to its mean. The mean of the transition rate $j_\mathrm{coll}(\hat{\mathbf{k}}_1,\omega_1,\hat{\mathbf{k}}_1',\omega_1')$ is simply the value of the term in the collision integral (Eq. (\ref{eq:Cint})) corresponding to the transition process $(\hat{\mathbf{k}}_1,\omega_1)\rightarrow(\hat{\mathbf{k}}_1',\omega_1')$. We therefore obtain the following autocorrelation function for the noise source:
\begin{eqnarray}
&&\langle \delta j(\hat{\mathbf{k}}_1,\omega_1,\mathbf{r}_1,t_1) j(\hat{\mathbf{k}}_2,\omega_2,\mathbf{r}_2,t_2)\rangle = \int\frac{d\hat{\mathbf{k}}'_1}{2\pi}\int\frac{d\omega'_1}{2\pi}\int\frac{d\hat{\mathbf{k}}'_2}{2\pi}\int\frac{d\omega'_2}{2\pi} \nonumber \\
&&\Bigg[\langle\delta j(\hat{\mathbf{k}}'_1,\omega'_1,\hat{\mathbf{k}}_1,\omega_1,\mathbf{r}_1,t_1)\delta j(\hat{\mathbf{k}}'_2,\omega'_2,\hat{\mathbf{k}}_2,\omega_2,\mathbf{r}_2,t_2)\rangle \nonumber \\
&&+\langle\delta j(\hat{\mathbf{k}}_1,\omega_1,\hat{\mathbf{k}}'_1,\omega'_1,\mathbf{r}_1,t_1)\delta j(\hat{\mathbf{k}}_2,\omega_2,\hat{\mathbf{k}}'_2,\omega'_2,\mathbf{r}_2,t_2)\rangle \nonumber \\
&&-\langle\delta j(\hat{\mathbf{k}}'_1,\omega'_1,\hat{\mathbf{k}}_1,\omega_1,\mathbf{r}_1,t_1)\delta j(\hat{\mathbf{k}}_2,\omega_2,\hat{\mathbf{k}}'_2,\omega'_2,\mathbf{r}_2,t_2)\rangle \nonumber \\
&&-\langle\delta j(\hat{\mathbf{k}}_1,\omega_1,\hat{\mathbf{k}}'_1,\omega'_1,\mathbf{r}_1,t_1)\delta j(\hat{\mathbf{k}}'_2,\omega'_2,\hat{\mathbf{k}}_2,\omega_2,\mathbf{r}_2,t_2)\rangle\Bigg],
\end{eqnarray}
where 
\begin{eqnarray}
&&\langle \delta j(\hat{\mathbf{k}}_1,\omega_1,\hat{\mathbf{k}}'_1,\omega'_1,\mathbf{r}_1,t_1)\delta j(\hat{\mathbf{k}}_2,\omega_2,\hat{\mathbf{k}}'_2,\omega'_2,\mathbf{r}_2,t_2)\rangle = \delta(\mathbf{r}_1-\mathbf{r}_2)\delta(t_1-t_2) \nonumber\\
&&\times(2\pi)^4\delta(\hat{\mathbf{k}}_1-\hat{\mathbf{k}}_2)\delta(\omega_1-\omega_2)\delta(\hat{\mathbf{k}}'_1-\hat{\mathbf{k}}'_2)\delta(\omega'_1-\omega'_2)\bar{j}_\mathrm{coll}(\hat{\mathbf{k}}_1,\omega_1,\hat{\mathbf{k}}'_1,\omega'_1).
\end{eqnarray}
This yields
\begin{eqnarray}
&&\langle\delta j(\hat{\mathbf{k}}_1,\omega_1,\mathbf{r}_1,t_1)\delta j(\hat{\mathbf{k}}_2,\omega_2,\mathbf{r}_2,t_2)\rangle \nonumber \\
&&= \delta(\mathbf{r}_1-\mathbf{r}_2)\delta(t_1-t_2)J(\mathbf{r}_1,\hat{\mathbf{k}}_1,\omega_1,\hat{\mathbf{k}}_2,\omega_2) \nonumber \\
&&=\delta(\mathbf{r}_1-\mathbf{r}_2)\delta(t_1-t_2)\Bigg[(2\pi)^2\delta(\hat{\mathbf{k}}_1-\hat{\mathbf{k}}_2)\delta(\omega_1-\omega_2)\Big\{\bar{j}_\mathrm{in}(\hat{\mathbf{k}}_1,\omega_1)+\bar{j}_\mathrm{out}(\hat{\mathbf{k}}_1,\omega_1)\Big\} \nonumber \\
&&~~~~~~~~~~~~~-\bar{j}(\hat{\mathbf{k}}_1,\omega_1,\hat{\mathbf{k}}_2,\omega_2)-\bar{j}(\hat{\mathbf{k}}_2,\omega_2,\hat{\mathbf{k}}_1,\omega_1)\Bigg].
\label{eq:nscf}
\end{eqnarray}
The averages are
\begin{eqnarray}
&&\bar{j}_\mathrm{in}(\hat{\mathbf{k}}_1,\omega_1) = v^2\int\frac{d\hat{\mathbf{k}}'}{2\pi}(1-f(\hat{\mathbf{k}}_1,\omega_1))f(\hat{\mathbf{k}}',\omega_1) \nonumber \\
&&+ \frac{{g'}^2}{2}\int\frac{d\Omega}{2\pi}\int\frac{d\hat{\mathbf{k}}'}{2\pi}\int\frac{d\omega'}{2\pi}B_\mathrm{loc}(\Omega)\Bigg[2\pi\delta(\omega'-\omega_1-\Omega)\Big\{\left(n_B(\Omega)+1\right)\left(1-f(\hat{\mathbf{k}}_1,\omega_1)\right) \nonumber \\
&&\times f(\hat{\mathbf{k}}',\omega')\Big\}+ 2\pi\delta(\omega'-\omega_1+\Omega)\Big\{n_B(\Omega)\left(1-f(\hat{\mathbf{k}}_1,\omega_1)\right)  f(\hat{\mathbf{k}}',\omega')\Big\} \Bigg],
\label{eq:noisemeans1}
\end{eqnarray}
\begin{eqnarray}
&&\bar{j}_\mathrm{out}(\hat{\mathbf{k}}_1,\omega_1) = v^2\int\frac{d\hat{\mathbf{k}}'}{2\pi}(1-f(\hat{\mathbf{k}}',\omega_1))f(\hat{\mathbf{k}}_1,\omega_1) \nonumber \\
&&+\frac{{g'}^2}{2}\int\frac{d\Omega}{2\pi}\int\frac{d\hat{\mathbf{k}}'}{2\pi}\int\frac{d\omega'}{2\pi}B_\mathrm{loc}(\Omega)\Bigg[2\pi\delta(\omega'-\omega_1-\Omega)\Big\{n_B(\Omega) \left(1-f(\hat{\mathbf{k}}',\omega')\right) \nonumber \\
&&\times f(\hat{\mathbf{k}}_1,\omega_1)\Big\} + 2\pi\delta(\omega'-\omega_1+\Omega)\Big\{\left(n_B(\Omega)+1\right) \left(1-f(\hat{\mathbf{k}}',\omega')\right) f(\hat{\mathbf{k}}_1,\omega_1)\Big\} \Bigg],
\label{eq:noisemeans2}
\end{eqnarray}
\begin{eqnarray}
&&\bar{j}(\hat{\mathbf{k}}_1,\omega_1,\hat{\mathbf{k}}_2,\omega_2) = 2 \pi v^2\delta(\omega_1-\omega_2)(1-f(\hat{\mathbf{k}}_2,\omega_1))f(\hat{\mathbf{k}}_1,\omega_1) \nonumber \\
&&+\frac{{g'}^2}{2}\int\frac{d\Omega}{2\pi}B_\mathrm{loc}(\Omega)\Bigg[2\pi\delta(\omega_2-\omega_1-\Omega)\Big\{n_B(\Omega) \left(1-f(\hat{\mathbf{k}}_2,\omega_2)\right) f(\hat{\mathbf{k}}_1,\omega_1)\Big\} \nonumber \\
&&+2\pi\delta(\omega_2-\omega_1+\Omega)\Big\{\left(n_B(\Omega)+1\right) \left(1-f(\hat{\mathbf{k}}_2,\omega_2)\right) f(\hat{\mathbf{k}}_1,\omega_1)\Big\} \Bigg].
\label{eq:noisemeans3}
\end{eqnarray}

To determine the noise, $f(\hat{\mathbf{k}}, \omega)$ in Eqs. (\ref{eq:noisemeans1}-\ref{eq:noisemeans3}) is set to the solution of the Boltzmann equation without noise terms, but with electromagnetic fields, as the noise is a weak perturbation. Once again, it should be noted that the $\mathbf{r},t$ dependence in Eqs. (\ref{eq:noisemeans1}-\ref{eq:noisemeans3}) is suppressed because it is the same for all terms.

\subsection{Computing the Fano Factor}

The noise is given by the correlation function of the current obtained from the Boltzmann equation, {\it i.e.}
\begin{equation}
S = \int dt\int d^2\mathbf{r}~\langle\mathbf{j}(\mathbf{r},t)\cdot\mathbf{j}(0,0)\rangle.
\end{equation}
This is not to be confused with the usual current correlation function in the Kubo formula; the current in the Boltzmann equation is an average current and is not the microscopic current operator of the Hamiltonian. The current density is given by
\begin{equation}
\mathbf{j}(\mathbf{r},t) = v_F\int\frac{d\hat{\mathbf{k}}}{2\pi}\int\frac{d\omega}{2\pi}\hat{\mathbf{k}}f(\hat{\mathbf{k}},\omega,\mathbf{r},t).
\label{eq:current}
\end{equation}
The Fano factor $F$ is given by $F=S/(2I)$ when the applied potential energy difference across the conductor is much larger than temperature, where $I=|(\int d^2\mathbf{r}~\mathbf{j})/\int d^2\mathbf{r}|$ is the current flowing through the conductor. The general relationship is given by
\begin{equation}
S = 2IF\coth\left(\frac{V}{2T}\right) + 4T(1-F)\frac{dI}{dV},
\end{equation}
where $V$ is the voltage applied across the conductor. We use this relation to compute
\begin{equation}
F = \frac{S-4T\frac{dI}{dV}}{2\left[I\coth\left(\frac{V}{2T}\right)-2T\right]\frac{dI}{dV}}.
\label{eq:FFS}
\end{equation}
Since $S\rightarrow =4T(dI/dV)$ as $V\rightarrow0$ (the Johnson-Nyquist thermal noise guaranteed by the fluctuation-dissipation theorem), the above equation subtracts off this contribution while computing the Fano factor. This aligns with the procedure used in experiments such as that of Ref. \cite{Natelson23}. 

We apply an electric field $\mathbf{E}$ to the system by transforming $\mathbf{k}\rightarrow\mathbf{k}-\mathbf{A}$; where $\mathbf{E}=-d\mathbf{A}/dt$. The steady-state Boltzmann equation in the absence of the noise source, which we need to solve in order to determine the $f$ from which the noise corrections are obtained \cite{deJongthesis,Kogan69}, becomes
\begin{eqnarray}
&&v_F\hat{\mathbf{k}}\cdot\mathbf{E}\frac{\partial f}{\partial \omega}+v_F\hat{\mathbf{k}} \cdot\frac{\partial f}{\partial\mathbf{r}} = I_{\mathrm{coll}}.
\label{eq:BE_driven}
\end{eqnarray}
Here we have assumed that $\partial\mathrm{Re}\Sigma^R/\partial \mathbf{r}$ and $\partial\mathrm{Re}\Sigma^R/\partial \mathbf{k}$ remain zero when $\mathbf{E}$ is applied, which follows from similar arguments to the ones we made for the case of the noise source; such an assumption should be reasonable in the $v-g'$ model as long as the potential energy difference across the conductor is much smaller than the Fermi energy. We also assume that the boson distribution function is unaffected by $\mathbf{E}$, which we justify as follows: the fermion distribution function $f$, and hence the Keldysh fermion Green's function $G^K$, are changed by an amount proportional to $\hat{\mathbf{k}}\cdot\mathbf{E}$ \cite{Patel:2017mjv, NaveLee07}. The boson self energy depends only upon the local fermion Green's functions ({\it i.e.} the Green's functions integrated over momentum) \cite{Patel21}. The change in these quantities is zero because $\int \hat{d\mathbf{k}}~\hat{\mathbf{k}}\cdot\mathbf{E} = 0$. Therefore the boson self energy (and hence the boson Green's functions and distribution function) are unaffected by the application of $\mathbf{E}$.

To solve this equation in the 1D conductor geometry required for noise computations, we use the ansatz \cite{deJongpaper}
\begin{equation}
f(\hat{\mathbf{k}},\omega,\mathbf{r}) = n_F(\omega) + n_F'(\omega)(\phi(\mathbf{r}=x\hat{\mathbf{x}})-V u(\hat{\mathbf{k}},\omega,\mathbf{r}=x\hat{\mathbf{x}})),
\label{eq:fansatz}
\end{equation}
and take $\mathbf{E}=E\hat{\mathbf{x}}$. Here, $\phi(\mathbf{r}=x\hat{\mathbf{x}}) = V-Ex$ is the electric potential. The conductor ranges from $x=0$ to $x=L$, and $V=-EL$. The boundary conditions to use for $u$ are \cite{deJongpaper}
\begin{equation}
u(\hat{\mathbf{k}},\omega,0) = 1,~~\hat{\mathbf{k}} \in \left[0,\frac{\pi}{2}\right]; ~~u(\hat{\mathbf{k}},\omega,L) = 0,~~\hat{\mathbf{k}} \in \left[\frac{\pi}{2},\pi\right].
\label{uboundary}
\end{equation}

We plug the parametrization Eq. (\ref{eq:fansatz}) into Eq. (\ref{eq:BE_driven}) to get (to linear order in $\mathbf{E}$)
\begin{eqnarray}
&&-V v_F k_x n_F'(\omega)\frac{\partial u(\hat{\mathbf{k}},\omega,x)}{\partial x} =-V v^2 n_F'(\omega) \left[\int\frac{d\hat{\mathbf{k}}'}{2\pi}u(\hat{\mathbf{k}}',\omega,x)-u(\hat{\mathbf{k}},\omega,x)\right] \nonumber \\
&&+ \delta I^{g'}_{coll}[n'_f(\omega)(\phi(x)- V u(\hat{\mathbf{k}},\omega,x))]
\end{eqnarray}

It is useful to define 
\begin{align}
    u(\phi,\omega,x)=1-\mathcal{T}(\pi-\phi, \omega,x)\,, \label{uT}
\end{align}
and we will see shortly that $\mathcal{T}$ has the interpretation of a transmission amplitude.
Then, the equation for this transmission amplitude becomes
\begin{equation}
\begin{split}
 -V v_F \cos \phi n'_F(\omega)\frac{\partial \mathcal{T}(\phi, \omega,x)}{\partial x} &=-V n'_F(\omega) v^2 \int\frac{d\phi'}{2\pi} \left[ \mathcal{T}(\phi,\omega,x) -\mathcal{T}(\phi',\omega,x)\right] +\\
 &+\delta I_\mathrm{coll}^{g'}[n'_F(\omega)( \Phi(x)-V)]+ \delta I_\mathrm{coll}^{g'}[n'_F(\omega)V \mathcal{T}(\phi,\omega,x)] .  
\end{split}
\label{eqT}
\end{equation}

\subsection{Solution for the transmission amplitude}
\label{sec:gp}

This subsection will exactly solve the differential equation for $\mathcal{T}$ in Eq.~(\ref{eqT}), subject to the boundary conditions in Eq.~(\ref{uboundary}).

But first, let us clarify the interpretation of $\mathcal{T}$ as a transmission amplitude. The linear differential equation in Eq.~(\ref{eq:BElin}) for the noise fluctuation $\delta f$ can be formally integrated using a Green's function
\begin{equation}
    \delta f(\hat{\mathbf{k}},\omega,\mathbf{r},t)=\int_{-\infty}^t d\omega' d\mathbf{r}' d\hat{\mathbf{k}}' dt' G(\hat{\mathbf{k}},\hat{\mathbf{k}}',\omega,\omega',\mathbf{r},\mathbf{r}',t-t')\delta j(\hat{\mathbf{k}}',\omega',\mathbf{r}',t') .\\
    \label{eqn:df_G}
\end{equation}
At finite temperature Eq. (\ref{eqn:df_G}) should be modified to include fluctuations at the boundary, but we show that those corrections are negligible, see Appendix \ref{app:FTC} for details.
Using Eq. (\ref{eq:current}) for the current density and Eq. (\ref{eq:noise}) for the noise power, we express the noise power in terms of the noise correlation function
\begin{equation} 
 S= \frac{2}{(2\pi)^4} \int d\mathbf{r}\int d\omega d\omega' d\hat{\mathbf{k}} d\hat{\mathbf{k}}'  \mathcal{T}(\hat{\mathbf{k}},\omega,\mathbf{r})\mathcal{T}(\hat{\mathbf{k}}',\omega',\mathbf{r})J(\mathbf{r},\hat{\mathbf{k}},\omega,\hat{\mathbf{k}}',\omega'),
 \label{eqn:shot_noise}
\end{equation}
where $J$ was defined in the second line of (\ref{eq:nscf}). The transmission amplitude $\mathcal{T}(\hat{\mathbf{k}},\omega,\mathbf{r})$ is expressed in terms of the Green's function
\begin{equation}
 \mathcal{T}(\hat{\mathbf{k}},\omega,\mathbf{r})=v_F\int_0^\infty dt' d\hat{\mathbf{k}}' k'_x d\omega'  dy'  G(\hat{\mathbf{k}}',\hat{\mathbf{k}},\omega',\omega,\mathbf{r}',\mathbf{r},t').
\end{equation}
Below $\mathbf{r}'$ is chosen to be at the right boundary of the sample, and $y'$ is integration over the width of the sample. The quantity $ \mathcal{T}(\hat{\mathbf{k}},\omega,\mathbf{r})$ is the same as that appearing in Eq.~(\ref{uT}), and can now be interpreted as the total probability of the particles with momenta $\hat{\mathbf{k}}$ and energy $\omega$ at position $\mathbf{r}$ to transfer to the end of the sample. 

\begin{figure}[h]
\begin{minipage}[h]{0.48\linewidth}
\center{\includegraphics[width=1\linewidth]{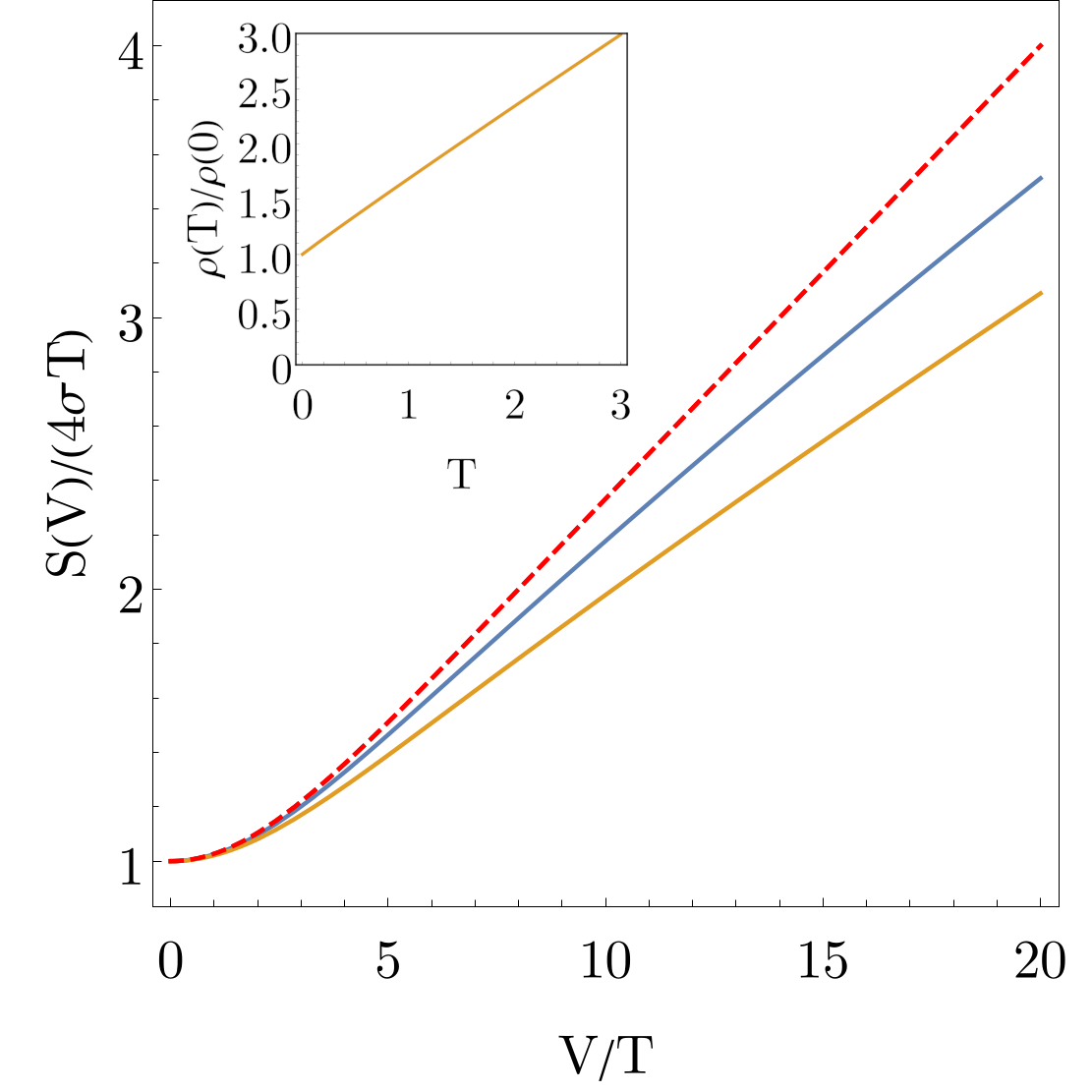}}
\\(a)
\end{minipage} 
\hfill    
\begin{minipage}[h]{0.48\linewidth}
\center{\includegraphics[width=1\linewidth]{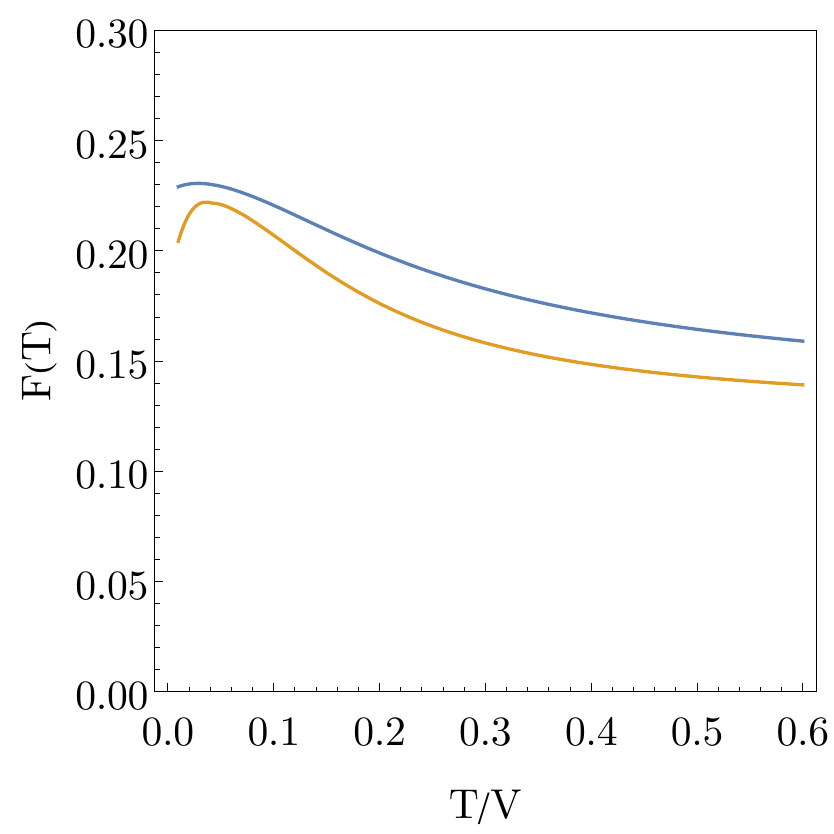}}
 \\(b)
\end{minipage} 
\caption{(a) Shot noise power as a function of the voltage $V$ at fixed temperature $T$. The red dashed line depicts shot noise power for a diffusive metal ($g'=0$), given by Eq. (\ref{eqn:s_diffusive}), while the blue and orange lines are the numerical solution for $g'>0$. $S(V=0)$ corresponds to $4 \sigma(g',T) T$. The inset shows $T$-linear resistivity for the same values of parameters as the orange line of the main plot. Parameters: $(k_F/v_F)^2 g'^2 T=g'^2 T/v^2=0.5,2$ (blue, orange) respectively. The shot noise power is measured in the units of $ W/L$.\
(b) Fano factor as a function of temperature $T$ at fixed voltage $V$. $F(T=0)$ is given in Eq. (\ref{eqn:fano_factor}), while the blue and orange lines are numerical solutions. Parameters: $(k_F/v_F)^2 g'^2 V=g'^2 V/v^2=45,320$ (blue, orange) respectively.}
\label{fig:shot_noise_np}
\end{figure}

Turning to the solution of Eq.~(\ref{eqT}), we search for solutions using the following ansatz
\begin{equation}
 \mathcal{T}(\phi,\omega,x)=\frac{x}{L}+A(\omega) \cos \phi.
\end{equation}
The $x$ dependence cancels out in Eq.~(\ref{eqT}), and we end up with
\begin{equation}
 -V v_F \cos \phi \frac{n'_F(\omega)}{L} =-V n'_F(\omega) v^2 A(\omega) \cos \phi+\delta I_\mathrm{coll}^{g'}[n'_F(\omega)V\cos \phi A(\omega)].
\end{equation}
The collision integral can be expressed in terms of $s(\omega)$, see Eq. (\ref{eqn:s_omega}). After that, we end up with a linear equation for $A(\omega)$
\begin{equation}
 - \frac{ v_F \cos \phi}{L} =- v^2 A(\omega) \cos \phi+A(\omega)\frac{\pi g'^2 v^2 s(\omega) \cos \phi}{v_F},
\end{equation}
 which suggests an answer
\begin{equation}
\mathcal{T}(\phi,\omega,x)=\frac{x}{L}+\frac{v_F/v^2}{L} \frac{\cos \phi}{1-\pi g'^2 s(\omega)/v_F}.
\label{eq:Tansatz}
\end{equation}
The solution has a simple interpretation: $l=v_F/v^2$ is a mean-free path, while non-zero $g'$ changes  its value. Note, that the following solution does not satisfy boundary conditions for short samples when $L \sim l$, but in the limit $L \gg l$ it becomes exact. Using the expression for the current density in Eq. (\ref{eq:current}) we compute the conductivity
\begin{equation}
 \sigma(g',T)=-L v_F \int \frac{\hat{\mathbf{k}}_x d\hat{\mathbf{k}}_x d\omega}{(2\pi)^2}\frac{\partial n_F(\omega)}{\partial \omega} \mathcal{T}(\hat{\mathbf{k}},x,\omega)=-\sigma_r \int d\omega \frac{n'_F(\omega)}{1-\pi g'^2 s(\omega)/v_F}. \label{sigmagp}
\end{equation}
This conductivity computed from the Boltzmann equation is exactly the same as that obtained from the Kubo formula using the fermion self energy in Eq. (\ref{eq:fse}) \cite{Aldape20, Patel21, Patel:2022gdh}, and gives rise to a linear $T$ dependence of the resistivity, which is shown in the inset of Fig. \ref{fig:shot_noise_np} (a). 

The transmission amplitude provides the following expression for the distribution inside the sample beyond linear order in $V$:
\begin{equation}
f(\hat{\mathbf{k}},\omega, x ,t)=(1-\mathcal{T}(-\hat{\mathbf{k}}, \omega, x))f_L(\omega)+\mathcal{T}(-\hat{\mathbf{k}},\omega, x) f_R(\omega),
\label{eqn:f}
\end{equation}
where $f_{L,R} (\omega)$ were defined in (\ref{fLR}). To obtain the noise power, we need to derive an expression for the noise correlation function, see Eq. (\ref{eq:nscf}).
We start with the $v$ term
\begin{eqnarray}
&&J^v(x,\hat{\mathbf{k}},\omega,\hat{\mathbf{k}}'
,\omega')=(2\pi)^2v^2 \delta(\hat{\mathbf{k}}-\hat{\mathbf{k}}') \delta(\omega-\omega')\times \nonumber \\
&&\int \frac{d\hat{\mathbf{k}}''}{2\pi}\left( (1-f(\hat{\mathbf{k}}))f(\hat{\mathbf{k}}'')+(1-f(\hat{\mathbf{k}}''))f(\hat{\mathbf{k}})\right) \nonumber \\
&&-v^2(2 \pi)\delta(\omega-\omega') \left( (1-f(\hat{\mathbf{k}}))f(\hat{\mathbf{k}}')+(1-f(\hat{\mathbf{k}}'))f(\hat{\mathbf{k}}) \right) ,  
\end{eqnarray}
where we omit the $\omega$ and $x$ dependence in $ f(\hat{\mathbf{k}},\omega, x)$. Substituting the expression above in Eq. (\ref{eqn:shot_noise}), we obtain
\begin{eqnarray}
&& S^v= 2W\frac{v^2}{(2\pi)^3}\int d\omega  d\hat{\mathbf{k}} d\hat{\mathbf{k}}' dx  (\mathcal{T}-\mathcal{T}')^2\left( (1-\mathcal{T})f_L(\omega)(1-f_L(\omega)) + \right. \nonumber \\
 &&\left.+\mathcal{T} f_R(\omega)(1-f_R(\omega))+(1-\mathcal{T}')\mathcal{T}(f_L(\omega)-f_R(\omega))^2\right),
\end{eqnarray}
where to lighten the notation we denote $\mathcal{T}=\mathcal{T}(\hat{\mathbf{k}},\omega,x)$ and $\mathcal{T}'=\mathcal{T}(\hat{\mathbf{k}}',\omega,x)$ and integration over $y$ gives the width of the sample $W$. We note, that it is enough to insert $\mathcal{T}=\mathcal{T}'=x/L$ in the second parenthesis because the momentum-dependent part will be of order $O(l/L)$. Therefore,  we end up with
\begin{eqnarray}
&&    S^v= 2\frac{v_F^2 W}{(2\pi)^3v^2 L^2}\int d\omega  d\hat{\mathbf{k}} d\hat{\mathbf{k}}' dx \frac{(k_x-k'_x)^2}{(1- \pi g'^2 s(\omega)/v_F)^2} \left( \left(1-\frac{x}{L} \right)f_L(\omega)(1-f_L(\omega)) + \right. \nonumber \\
 &&\left.+\frac{x}{L} f_R(\omega)(1-f_R(\omega))+\left(1-\frac{x}{L}\right)\frac{x}{L}(f_L(\omega)-f_R(\omega))^2\right).
\end{eqnarray}
Integration over momentum can be easily performed and the first two terms give
\begin{equation}
 S_0^v= 2 W\frac{v_F^2 }{(2\pi)L^2 v^2}\int \frac{d\omega  dx}{(1-\pi g'^2 s(\omega)/v_F)^2}    \left( \left(1-\frac{x}{L} \right) \left(-T\frac{\partial f_L}{\partial \omega} \right) +\frac{x}{L} \left(-T\frac{\partial f_R}{\partial \omega} \right)\right).
\end{equation}

After integrating over $x$
\begin{equation}
  S_0^v=-\frac{2 \sigma_r W T}{L}  \int d\omega \frac{\partial f_L/\partial \omega+\partial f_R/\partial \omega}{(1- \pi g'^2 s(\omega)/v_F)^2}.
  \label{eqn:v0}
\end{equation}
Similarly, the last term is 
\begin{equation}
 S_1^v=\frac{2 \sigma_r W}{3L}  \int d \omega \frac{(f_L(\omega)-f_R(\omega))^2}{(1- \pi g'^2 s(\omega)/v_F)^2}.
 \label{eqn:v1}
\end{equation}
Now we focus on $J^{g'}$ contribution to the shot noise power. Using Eq. (\ref{eqn:shot_noise})
\begin{eqnarray}
 &&S^{g'}=\frac{2\pi g'^2}{(2\pi)^5} W \int dx d\omega d\omega' d\hat{\mathbf{k}} d\hat{\mathbf{k}}' B_\mathrm{loc}(\omega'-\omega)(\mathcal{T}(\hat{\mathbf{k}},\omega,x)-\mathcal{T}(\hat{\mathbf{k}}',\omega',x))^2\times \nonumber \\
 &&\left[(n_B(\omega'-\omega)+1)(1-f)f'+n_B(\omega'-\omega)(1-f')f\right],    
\end{eqnarray}
where again to lighten the notation we introduce $f=f(\hat{\mathbf{k}},\omega, x)$ and $f'=f(\hat{\mathbf{k}}',\omega', x)$.
As in the previous calculation we account for $\hat{\mathbf{k}}$ dependence only in the term $(\mathcal{T}-\mathcal{T}')$ where the $x/L$ part cancels. This allows us to perform integration over momentum
\begin{eqnarray}
     &&S^{g'}=\frac{\pi g'^2 v_F^2}{(2\pi)^3 L^2 v^4} W \int dx d\omega d\omega'  \left[ \frac{ B_\mathrm{loc}(\omega'-\omega)}{(1-\pi g'^2 s(\omega)/v_F)^2}+\frac{ B_\mathrm{loc}(\omega'-\omega)}{(1- \pi g'^2 s(\omega')/v_F)^2} \right] \nonumber \\
     &&\times \left[(n_B(\omega'-\omega)+1)(1-f)f'+n_B(\omega'-\omega)(1-f')f\right].
\end{eqnarray}

The integration over $x$ can be performed straightforwardly and we end up with
\begin{eqnarray}
   && S^{g'}=\frac{4 \pi g'^2 \sigma_r}{(2\pi)^2 L v^2} W \int  d\omega d\omega'  \left[ \frac{ B_\mathrm{loc}(\omega'-\omega)}{(1-\pi g'^2 s(\omega)/v_F)^2}+\frac{ B_\mathrm{loc}(\omega'-\omega)}{(1-\pi g'^2 s(\omega')/v_F)^2} \right]n_B(\omega'-\omega) \nonumber \\
   && \times \left(\frac{f_L+f_R}{2}-\frac{f_L f'_L+f_R f'_R}{3}-\frac{f'_Lf_R+f_L f'_R}{6}\right) .
   \label{eqn:g'}
\end{eqnarray}

Combining all terms $S=S^v_0+S^v_1+S^{g'}$ we obtain Eq. (\ref{eq:mainresult}). 

Eq. (\ref{eq:mainresult}) can be further simplified in two limits. In the limit $T=0$ the local boson spectral function in Eq. (\ref{eqn:bloc}) becomes $B_{\mathrm{loc}}(\omega)=(1/4) \sgn (\omega)$ and
\begin{equation}
 s(\omega)=-\frac{v_F |\omega|}{2(2 \pi)^2 v^2}.
\end{equation}
The $S_0^v$ term vanishes and $S_1^v$ term after integration over frequency is
\begin{equation}
  S^v=2 \frac{ v_F^2 W}{6(2\pi) v^2 L} \frac{V}{1+\pi g'^2 V/(2 (2\pi)^2 v^2)}=\frac{2 V\sigma_r W}{3L}\frac{1}{1+\pi g'^2 V/(2 (2\pi)^2 v^2)}.
\end{equation}
The $S^{g'}$ term also simplifies
\begin{equation}
 S^{g'}=\frac{\pi g'^2 v_F^2}{12(2\pi)^3 L^2 v^4} W \int_0^V d\omega \int_\omega^V d\omega'  \left[ \frac{ 1}{(1-\pi g'^2 s(\omega)/v_F)^2}+\frac{ 1}{(1-\pi g'^2 s(\omega')/v_F)^2} \right].
\end{equation}
Integrations over frequencies can be performed exactly and we obtain
\begin{equation}
 S^{g'}=\frac{\pi g'^2 v_F^2}{12(2\pi)^3 L v^4} W \frac{V^2}{1+\pi g'^2 V/(2(2\pi)^2v^2)}=\frac{2\sigma_r W V}{3L}\frac{\pi g'^2 V}{4(2\pi)^2v^2} \frac{1}{1+\pi g'^2 V/(2(2\pi)^2v^2)}.
\end{equation}

Summing up both contributions
\begin{equation}
 S=S^v+S^{g'}=\frac{2\sigma_r W V}{3L} \frac{1+\pi g'^2 V/(4(2\pi)^2v^2)}{1+\pi g'^2 V/(2(2\pi)^2v^2)}.
\end{equation}
Now it is straightforward to compute the Fano factor, and we obtain the result in Eq.~(\ref{eqn:fano_factor}).

Another limit which can be analyzed is $V=0$. In this limit $S^v_1=0$ and 
\begin{equation}
 S_0^v=-\frac{4 \sigma_r W T}{L}  \int d\omega \frac{n'_F(\omega)}{(1- \pi g'^2 s(\omega)/v_F)^2}.
\end{equation}
$S^{g'}$ term can also be simplified
\begin{eqnarray}
  && S^{g'}=\frac{8 \pi g'^2 \sigma_r}{(2\pi)^2 L v^2} W \int  d\omega d\omega'  \left[ \frac{ B_\mathrm{loc}(\omega'-\omega)}{(1-\pi g'^2 s(\omega)/v_F)^2}+\frac{ B_\mathrm{loc}(\omega'-\omega)}{(1-\pi g'^2 s(\omega')/v_F)^2} \right]\times \nonumber \\
  &&\times n_B(\omega'-\omega)n_F(\omega)\left(1-n_F(\omega')\right)  .
\end{eqnarray}

After some algebra it is possible to demontrate that two terms in parenthesis are equal and
\begin{equation}
 S^{g'}=-2T\frac{2 g'^2 \sigma_r}{2\pi v^2 L } W \int  d\omega d\omega'  \left[ \frac{ B_\mathrm{loc}(\omega'-\omega)}{(1-\pi g'^2 s(\omega)/v_F)^2}\right]\left( n_B(\omega'-\omega)+n_F(\omega')\right)n'_F(\omega),
\end{equation}
which can be rewritten as

\begin{equation}
 S^{g'}=T\frac{4g'^2 \sigma_r}{v_F L } W \int  d\omega d\omega'   \frac{s(\omega)n'_F(\omega)}{(1-\pi g'^2 s(\omega)/v_F)^2} .
\end{equation}
Therefore, the fluctuation-dissipation theorem can be established, providing the Johnson-Nyquist thermal noise,
\begin{equation}
 S=S_0^v+S^{g'}=-\frac{4 \sigma_r W T}{L}  \int d\omega \frac{n'_F(\omega)(1- \pi g'^2 s(\omega)/v_F)}{(1- \pi g'^2 s(\omega)/v_F)^2}=\frac{4 \sigma(g',T) W T}{L} \,, \label{JNg}
\end{equation}
where $\sigma (g',T)$ is the conductivity obtained in (\ref{sigmagp}).

Fig. \ref{fig:shot_noise_np} (a) shows the shot noise power as a function of the voltage for both $g'=0$ and $g'>0$. $S(V=0)$ is precisely equal to $4 \sigma(g',T) T$, which is guaranteed by the fluctuation-dissipation theorem. The function's slope is also smaller at $g'>0$, resulting in a smaller Fano factor.
Fig. \ref{fig:shot_noise_np} (b) shows the Fano factor obtained using Eq. (\ref{eq:FFS}), where we take $I=\sigma(g',T)V$. There is an upturn as the function of temperature, but as temperature increases, it goes down. We note a similarity with experiment, though in the experiment the Fano factor appears to decay more strongly as a function of temperature. Note that the dimensionless couplings in the second plot are much higher, since we always need $g'^2 V/v^2 \gg 1$ to observe the reduced Fano factor.

In Appendix \ref{app:NC}, we present results for the Fano factor away from the QCP. In this regime, transport is Fermi liquid like at low $T$, with a $T^2$ resistivity (as opposed to the $T$-linear resistivity at the QCP). The weaker inelastic electron scattering consequently leads to a smaller suppression of the Fano factor relative to its non-interacting value of $F=1/3$.

\section{Discussion}
\label{sec:discussion}

We have presented a general framework for the computation of shot noise in metals without quasiparticles. We generalized the noisy Boltzmann equation formalism for disordered metals with quasiparticles \cite{Kogan69,deJongBeen,deJongthesis} to the situation without quasiparticles by allowing the electron distribution function to have independent dependence on the excitation energy $\omega$ and the momentum $\mathbf{k}$
\cite{PrangeKadanoff,NaveLee07,Patel:2017mjv}. 

We applied this framework to a recently developed theory of strange metals \cite{Aldape20,Patel21,Guo2022,Patel:2022gdh} with a spatially random Yukawa coupling $g' (\mathbf{r})$ between the electrons and a critical boson, along with a spatially random potential $v (\mathbf{r})$ on the electrons. A spatially uniform coupling $g$ cancels out in the linear-response transport properties to leading order \cite{Patel:2022gdh,Guo2022,Shi22}, and we have assumed that a similar cancellation is present in the non-linear transport response functions needed for shot-noise, as we will describe further below. The $g'$-$v$ theory led to differential equations which were similar to those obtained earlier \cite{Kogan69,deJongBeen,deJongthesis}, and could also be solved in a wire geometry in the presence of an applied voltage. 
Our main result in (\ref{eq:mainresult}) describes the dependence of the shot-noise on $g'$, $v$, the applied voltage, and temperature. In the limit of low temperature, it reduces the simple result in (\ref{eqn:fano_factor}), which shows the crossover from the noninteracting value of the Fano factor $F=1/3$ for ${g'}^2V \ll v^2$, to the strongly interacting non-quasiparticle value $F=1/6$ for ${g'}^2V \gg v^2$. Also interesting is the non-monotonic dependence of the Fano factor on temperature in Fig.~\ref{fig:shot_noise_np} (b), starting from the $T=0$ values in (\ref{eqn:fano_factor}); this is a prediction of our theory can be tested by extending existing measurements to lower $T$.

We now comment on the effects of a spatially uniform Yukawa coupling ($g$), which we ignored in our calculations. As shown in earlier work \cite{Patel:2022gdh,Guo2022}, the dominant part of the inelastic scattering of electrons induced by this coupling is forward scattering, which does not relax current and momentum. Hence, it is incapable of generating current fluctuations leading to noise \cite{Nagaev95}, and furthermore, it also does not relax the shift in the fermion distribution function induced by the applied voltage, which is in the same angular harmonic channel as the current. Put together, these two points ensure that both the conductivity and the noise are therefore not affected by this critical forward scattering. However, the $g$ coupling will also induce non-critical large angle scattering, mediated by effectively instantaneous interactions arising from gapped boson modes. Such processes are akin to the usual electron-electron scattering from screened Coulomb interactions in Fermi liquids; while they do not directly cause current fluctuations and noise as they still can't change the total momentum of the system and thereby relax current, they do affect the fermion distribution function in a way that has been shown to increase the Fano factor above $F=1/3$ by thermalizing the energy added by the applied voltage around the Fermi surface \cite{Nagaev95, Kozub95}.

In contrast, the current and momentum relaxing inelastic scattering from the spatially random $g'$ interactions considered in this work are capable of generating current fluctuations and noise. Additionally, they relax the voltage-induced shift in the fermion distribution function while dumping the added energy into the bosonic bath. As we have shown, the net result of these effects is a reduction in the Fano factor with respect to $F=1/3$; this is not unlike the case of inelastic electron-phonon scattering considered in Refs. \cite{Nagaev95, Kozub95}, which has similar relaxational properties. We therefore conclude that the observation of a Fano factor $F\ll 1/3$ as $T\rightarrow0$ in a quantum critical metal must imply the existence of a significant spatially random component in the critical electron-electron interactions which is capable of overwhelming the $F$-increasing effects of large-angle scattering from the spatially uniform non-critical component. In a sufficiently long sample, the non-critical large angle scatterings will eventually dominate and thermalize the electron distribution function to a local temperature; so observation of our predictions for the $g'$-$v$ model requires short samples.     

In Appendix \ref{app:KLM} we extend our results to a Kondo lattice model with spatially random interactions \cite{Aldape20}.
Such a model is more appropriate for the recent experiment \cite{Natelson23} on a heavy fermion material than the single band model considered in the body of the paper. Here the boson $\phi$ is realized by the hybridization between the conduction electron and the fermionic spinons (the `slave' boson), and this boson becomes critical at the non-symmetry-breaking transition between small and large Fermi surfaces.
We show in Appendix~\ref{app:KLM} that the 
noise properties of this Kondo lattice model are essentially the same as those of the one-band model described in the body of the paper.

Given that quasi-elastic electron-phonon scattering can also give rise to $T$-linear resistivity for temperatures above a finite temperature scale \cite{ZimanBook}, and that such a mechanism is a strong contender for the origin of $T$-linear resistivity in moir\'e graphene bilayers except at the lowest temperature scales \cite{DasSarma2022}, it is interesting to ask whether shot noise measurements can differentiate between this classic mechanism of $T$-linear resistivity and the quantum critical mechanism considered in the bulk of this paper that involves mostly inelastic scattering. In Appendix \ref{app:HP}, we compute shot noise for electron-phonon scattering with dispersionless Holstein phonons. In the quasi-elastic regime, which is where $T$-linear resistivity occurs, we find that $F=1/3$, even when the $T$-linear piece of the resistivity is much larger than the residual impurity-induced resistivity. This is in marked contrast to our results for quantum criticality, where $F\ll 1/3$ in the $T$-linear dominated regime. We therefore propose shot noise measurements as a purely electrical tool that can shed further light on the origins of $T$-linear resistivity in various materials, with Fano factors significantly smaller than $1/3$ in the $T$-linear regime indicating the dominance of inelastic scattering in transport ({\it i.e.} current degrading inelastic scattering), and therefore true non-Fermi liquid behavior. 

\acknowledgements

We thank Alex Levchenko, Doug Natelson, Qimiao Si, and Yiming Wang for valuable discussions, and Doug Natelson for informing us about the observations in Ref.~\cite{Natelson23}.
This research was supported by the U.S. National Science Foundation grant No. DMR-2245246, and by the Simons Collaboration on Ultra-Quantum Matter which is a grant from the Simons Foundation (651440, S.S.). The Flatiron Institute is a division of the Simons Foundation.

\appendix

\section{Finite temperature boundary corrections}
\label{app:FTC}

Strictly speaking, Eq. (\ref{eqn:df_G}) should be modified at non-zero temperature to include fluctuations coming from left and right reservoirs, see Ref.~\cite{deJongthesis}. 
\begin{eqnarray}
&&\delta f(\hat{\mathbf{k}},\omega,\mathbf{r},t)=\int_{-\infty}^t d\mathbf{r}' d\hat{\mathbf{k}}' dt' G(\hat{\mathbf{k}},\hat{\mathbf{k}}',\mathbf{r},\mathbf{r}',t-t')\delta j(\hat{\mathbf{k}}',\omega,\mathbf{r}',t')+\nonumber \\
&&+\int_{-\infty}^t dt' \int_{S_L}d\mathbf{r}'\int_{k_x>0} d\hat{\mathbf{k}}'  v_F k_x G(\hat{\mathbf{k}},\hat{\mathbf{k}}',\mathbf{r},\mathbf{r}',t-t')\delta f(\hat{\mathbf{k}}',\omega,\mathbf{r}',t')+\mathrm{(R.R.)},
\end{eqnarray}
where $S_L$ indicates integration over the left boundary of the sample and there is a corresponding term at the right boundary denoted as (R.R.).
The correlation function in these reservoirs is given by
\begin{equation}
 \langle  \delta f(\hat{\mathbf{k}},\omega,\mathbf{r},t)   \delta f(\hat{\mathbf{k}}',\omega',\mathbf{r}',t') \rangle=(2\pi)^2\delta(\omega-\omega')\delta(\hat{\mathbf{k}}-\hat{\mathbf{k}}')\delta(\mathbf{r}-\mathbf{r}'-v \hat{\mathbf{x}}(t-t'))f_{L,R}(\omega)(1-f_{L,R}(\omega)).
\end{equation}
The Eq. (\ref{eqn:shot_noise}) for the shot noise power is modified by the following boundary terms

\begin{eqnarray}
&& S^b= \frac{2 v_F}{(2\pi)^2} \int d\omega  d\hat{\mathbf{k}} k_x \mathcal{T}(\hat{\mathbf{k}},\omega,0)^2f_L(\omega)(1-f_L(\omega))+\nonumber \\
&&+\frac{2 v_F}{(2\pi)^2} \int d\omega  d\hat{\mathbf{k}} k_x \left(1-\mathcal{T}(\hat{\mathbf{k}},\omega,L) \right)^2f_R(\omega)(1-f_R(\omega)) . 
\end{eqnarray}
One might notice that those terms contain additional powers of $l/L$ compared to $S^v$ and $S^{g'}$, and should therefore not be included. Moreover, $\mathcal{T}(\hat{\mathbf{k}},\omega,0)=\mathcal{T}(\hat{\mathbf{k}},\omega,L)-1 \sim k_x^2$, and the boundary terms must vanish after integration over $\hat{\mathbf{k}}$.

\section{Shot noise away from criticality}
\label{app:NC}
\begin{figure}[h]
\begin{minipage}[h]{0.45\linewidth}
\center{\includegraphics[width=1\linewidth]{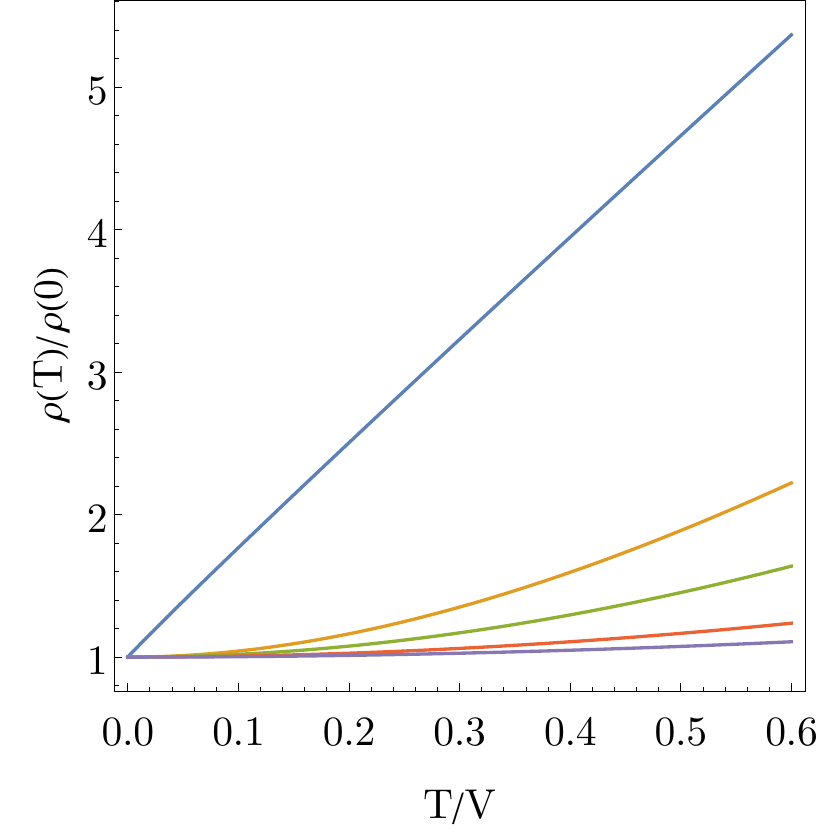}}
\\(a)
\end{minipage} 
\hfill  
\begin{minipage}[h]{0.45\linewidth}
\center{\includegraphics[width=1\linewidth]{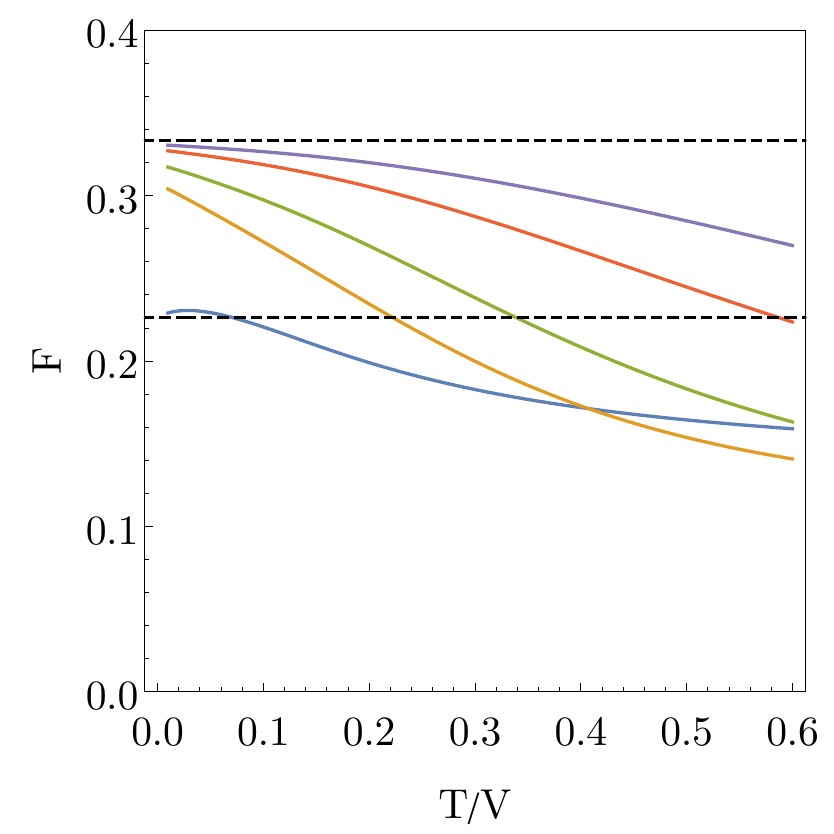}}
\\(b)
\end{minipage} 
\caption{(a) Resistivity as a function of temperature for $\Delta \kappa=0~(\mathrm{critical}),-2,-4,-10,-20$ moving away from criticality. Lines are shown from top to bottom.
(b) Corresponding Fano factors as a function of temperature $T$ at fixed voltage $V$, with the same color coding. The dashed lines are given by Eq. (\ref{eqn:fano_factor}) and $F=1/3$. Parameters: $(k_F/v_F)^2 g'^2 V=g'^2 V/v^2=45$ respectively.}
\label{fig:shot_noise_m}
\end{figure}

We can address the question of what happens away from criticality by solving the equation Eq. (\ref{eq:thermal_mass}) for finite $\Delta\kappa<0$. The resistivity is linear with temperature in the critical regime, while it switches to quadratic behaviour away from criticality, see Fig. \ref{fig:shot_noise_m}(a). 
At $T=0$ it is possible to get an analytic expression for $s(w)$;
\begin{eqnarray}
&&s(\omega)=-\frac{2 v_F}{(2\pi)^2 v^2}\int_0^{|\omega|}B_{loc}(\omega') d\omega' \nonumber \\
&&=-\frac{2 v_F}{(2\pi)^2 v^2}\left[ \frac{\omega}{2 \pi }\arctan \left(\frac{c_d \omega}{m^2(T=0)} \right)+\frac{m^2(T=0)}{c_d}\ln \left(1+\frac{c_d^2 \omega^2}{m^4(T=0)} \right)\right],
\end{eqnarray}
which allows for a simplified expression of Fano factor to be obtained:
\begin{equation}
 \frac{L}{2 W \sigma}\frac{dS}{dV}=\frac{1}{6}\frac{2-g'^2 s(V)/v_F}{(1-g'^2 s(V)/v_F)^2}+\frac{g'^2}{3(2\pi)^2 v^2}\int_{0}^V d\omega' \left[\frac{B_{loc}(\omega
 ')}{(1-g'^2 s(V-\omega')/v_F)^2} \right].
\end{equation}
When we are far from criticality and $m^2(T=0)>c_d V$, the Fano factor returns to its noninteracting value of $1/3$. This is not too surprising, because the finite boson mass at $T=0$ suppresses inelastic scattering at low energy. We can also analyze finite temperature scenarios by computing the integrals numerically, as shown in Fig. \ref{fig:shot_noise_m}. The Fano factor $F(T)$ increases as we move away from criticality at low $T$. 

\section{Kondo lattice model}
\label{app:KLM}

In this Appendix we will show that the computation of shot noise in a two-band Kondo lattice model \cite{Coleman1984, Senthil2004} augmented with $g'$ interactions \cite{Aldape20}, which is a more accurate description of heavy fermion compounds than the simple model of the main text, is equivalent to the computation in the main text under reasonable assumptions. 

The Kondo lattice model has two species of fermions ($c, f$), with the boson $\phi$ mediating hybridization between them. 
\begin{align}
\mathcal{L}_c & = \sum_{\mathbf{k},\sigma=\uparrow,\downarrow} c_{\mathbf{k},\sigma}^\dagger (\tau) \left( \frac{\partial}{\partial \tau} + \epsilon_{\mathbf{k}, c} \right) c_\mathbf{k,\sigma} (\tau) \nonumber \\
\mathcal{L}_f & = \sum_{\mathbf{k},\sigma} f_{\mathbf{k},\sigma}^\dagger (\tau) \left( \frac{\partial}{\partial \tau} + \epsilon_{\mathbf{k}, f} \right) f_\mathbf{k,\sigma} (\tau) \nonumber \\
\mathcal{L}_\phi & = \frac{1}{2} \int d^2 \mathbf{r} \, \left[\phi^{\ast}(\mathbf{r},\tau)\partial_\tau\phi(\mathbf{r}, \tau) +  |\mathbf{\nabla}_\mathbf{r} \phi (\mathbf{r}, \tau)|^2  + m_b^2 |\phi (\mathbf{r}, \tau)|^2 + \ldots \right] \nonumber \\
\mathcal{L}_{v} & =  \int d^2 \mathbf{r}\sum_{\sigma} \, \left[v_c(\mathbf{r}) c^\dagger_{\sigma} (\mathbf{r},\tau) c_{\sigma}  (\mathbf{r},\tau) + v_f(\mathbf{r}) f^\dagger_{\sigma} (\mathbf{r},\tau) f_{\sigma} (\mathbf{r},\tau)\right]  \nonumber \\
\mathcal{L}_{g-g'} & =  \int d^2 \mathbf{r}\sum_{\sigma} \, (g + g'(\mathbf{r})) \, c^{\dagger}_{\sigma}(\mathbf{r},\tau)  f_{\sigma}(\mathbf{r},\tau) \, \phi(\mathbf{r},\tau)\ + \mathrm{H.c.} 
\label{klm}
\end{align}
Here, $v_{c,f}(\mathbf{r})$ are mutually uncorrelated random potentials for $c,f$, and $g'(\mathbf{r})$ is the spatially random part of the Yukawa coupling. Since the $f$ fermions are spinons, there is an additional constraint 
\begin{equation}
\sum_{\sigma}f^\dagger_{\sigma}(\mathbf{r},\tau)f_{\sigma}(\mathbf{r},\tau) - \phi^{\ast}(\mathbf{r},\tau)\phi(\mathbf{r},\tau)=1,
\label{eq:constraint}
\end{equation}
that ensures a single valence spin per site. The constraint can be implemented by a Lagrange multiplier
acting as the time component of an emergent U(1) gauge field. For the calculation in this paper, this constraint is important for generating the boson thermal `mass' $m^2(T)$ (Eq. (\ref{eq:thermal_mass})), at finite temperature above the QCP \cite{Aldape20}.

While we have included the spatially independent part of the Yukawa coupling ($g$) in Eq. (\ref{klm}), its effects are unimportant at the QCP. Since the critical bosons are long wavelength ($\mathbf{q}\approx0$) excitations, and the $c$ and $f$ Fermi surfaces are generically not matched, the resulting large difference between $\epsilon_{\mathbf{k+q},c}$ and $\epsilon_{\mathbf{k},f}$ makes the transition rates induced by the $g$ coupling small. We can therefore drop it and retain only the spatially random part $g'(\mathbf{r})$ like in the simplified model in the main text.

Ref. \cite{Aldape20} showed that the quantum critical transport properties of the Kondo lattice model with $g'$ interactions are dominated by the $c$ fermions. The constraint (Eq. (\ref{eq:constraint})) implies that the conductivity is given by 
\begin{equation}
\sigma = \sigma_c + \frac{\sigma_f\sigma_b}{\sigma_f+\sigma_b}.
\label{eq:klsigma}
\end{equation}
The poor ($\mathcal{O}(e^2/h)$) conductivity $\sigma_b$ of the critical bosons in two spatial dimensions suppresses the contribution of the $f$ fermions $\sigma_f$ by series addition of resistivities, resulting in $\sigma\approx\sigma_c$, {\it i.e.} the conductivity of the conduction electrons. We will now show that the fluctuations in the current that lead to shot noise can also be computed in an effective one-band picture involving only the $c$ electrons, leading to a calculation that is equivalent to that in the main text.

Since the charge current is dominated by the $c$ electrons, we need to consider fluctuations only of this part of the current; the small current carried by the $f-\phi$ subsystem implies that its fluctuations are proportionately weaker and can be ignored. The Boltzmann equation for the $c$ electrons in the absence of a noise term is given by Eq. (\ref{eq:BE_orig}). However, the collision integrals are now slightly different \footnote{To derive Eq. (\ref{eq:Cintklm}) we have assumed that $B_{\mathrm{loc}}(\Omega)=-B_{\mathrm{loc}}(-\Omega)$, which is a feature of the model at strong coupling \cite{Aldape20}.}
\begin{eqnarray}
&& I_\mathrm{coll}^v = v_c^2\int\frac{d\hat{\mathbf{k}}'}{2\pi}\Bigg[\left(1-f_c(\hat{\mathbf{k}},\omega)\right)f_c(\hat{\mathbf{k}}',\omega)-\left(1-f_c(\hat{\mathbf{k}}',\omega)\right)f_c(\hat{\mathbf{k}},\omega)\Bigg], \nonumber \\
&& I_\mathrm{coll}^{g'} = \frac{{g'}^2}{2}\int\frac{d\Omega}{2\pi}\int\frac{d\hat{\mathbf{k}}'}{2\pi}\int\frac{d\omega'}{2\pi}B_\mathrm{loc}(\Omega)\Bigg[2\pi \delta(\omega'-\omega-\Omega)\Big\{\left(n_B(\Omega)+1\right) \nonumber \\
&&\times \left(1-f_c(\hat{\mathbf{k}},\omega)\right) f_f(\hat{\mathbf{k}}',\omega') - n_B(\Omega)\left(1-f_f(\hat{\mathbf{k}}',\omega')\right)f_c(\hat{\mathbf{k}},\omega)\Big\} \nonumber \\
&&+2\pi \delta(\omega'-\omega+\Omega)\Big\{n_B(\Omega)\left(1-f_c(\hat{\mathbf{k}},\omega)\right)f_f(\hat{\mathbf{k}}',\omega') \nonumber \\
&&-\left(n_B(\Omega)+1\right)\left(1-f_f(\hat{\mathbf{k}}',\omega')\right)f_c(\hat{\mathbf{k}},\omega)\Big\}\Bigg].
\label{eq:Cintklm}
\end{eqnarray}
The occurence of an $f_f$ along with an $f_c$ in a scattering term (instead of two $f_c$'s) is the consequence of the boson mediating hybridization between the $c$ and $f$ bands. However, due to Eqs. (\ref{eq:fansatz}, \ref{uT}, \ref{eq:Tansatz}), it can easily be established that \footnote{The $c$ and $f$ fermions are both subject to the same voltage $V$, and therefore have the same form of Eqs. (\ref{eq:fansatz}, \ref{uT}, \ref{eq:Tansatz}). The rescaling of the $f,b$ conductivities from $\sigma_{f,b}$ to $\sigma_{f,b}\sigma_{b,f}^2/(\sigma_f+\sigma_b)^2$ respectively follows from the renormalization of the $f,b$ charges due to the emergent U(1) gauge field \cite{Aldape20}. The total conductivity (Eq. (\ref{eq:klsigma})) is the sum of the $c$ conductivity and the rescaled $f,b$ conductivities.}
\begin{equation}
\int\frac{d\hat{\mathbf{k}}'}{2\pi}f_{c,f}(\hat{\mathbf{k}}',\omega') = n_F(\omega') + 2V\frac{x}{L}n_F'(\omega') \equiv n_F(\omega',x).
\label{eq:reduction}
\end{equation}

The collision integrals of both the Kondo lattice model (Eq. (\ref{eq:Cintklm})), as well as the simplified model in the main text, then reduce to the same form, {\it i.e.}
\begin{eqnarray}
&& I_\mathrm{coll}^v = v_c^2(n_F(\omega,x)-f_c(\hat{\mathbf{k}},\omega)), \nonumber \\
&& I_\mathrm{coll}^{g'} = \frac{{g'}^2}{2}\int\frac{d\Omega}{2\pi}\int\frac{d\omega'}{2\pi}B_\mathrm{loc}(\Omega)\Bigg[2\pi\delta(\omega'-\omega-\Omega)\Big\{\left(n_B(\Omega)+1\right)\left(1-f_c(\hat{\mathbf{k}},\omega)\right) n_F(\omega',x) \nonumber \\
&&- n_B(\Omega)\left(1-n_F(\omega',x)\right)f_c(\hat{\mathbf{k}},\omega)\Big\} +2\pi\delta(\omega'-\omega+\Omega)\Big\{n_B(\Omega)\left(1-f_c(\hat{\mathbf{k}},\omega)\right)n_F(\omega',x) \nonumber \\ 
&&-\left(n_B(\Omega)+1\right)\left(1-n_F(\omega',x)\right)f_c(\hat{\mathbf{k}},\omega)\Big\}\Bigg].
\label{eq:Cintuniversal}
\end{eqnarray}
It was established in Ref. \cite{Aldape20} that the boson propagator (and hence $B_\mathrm{loc}(\Omega)$) takes the same form as the expression in the main text at strong coupling. Therefore, the solution of the Boltzmann equation for the $c$ distribution function $f_c(\hat{\mathbf{k}},\omega)$, and hence the conductivity $\sigma\approx\sigma_c$, is identical to the solution obtained in the main text \footnote{This equivalence of the $c$ conductivity in the simple one-band model and the two-band Kondo lattice model can also be obtained directly from the Kubo formula, as can be seen by comparing the calculations in Ref. \cite{Patel:2022gdh} and Ref. \cite{Aldape20}.}. The $f$ collision integral also follows the same pattern as the above, which ensures that Eq. (\ref{eq:reduction}) is self-consistent. 

We now establish equivalence of the autocorrelation function of the noise source $\delta j$ in the $c$ Boltzmann equation. The modification of Eqs. (\ref{eq:noisemeans1}, \ref{eq:noisemeans2}) follows the same pattern as the transformation of Eq. (\ref{eq:Cint}) into Eq. (\ref{eq:Cintklm}), {\it i.e.} the distribution function that is integrated over $\hat{\mathbf{k}}$ is replaced by $f_f$. Then the application of Eq. (\ref{eq:reduction}) causes Eqs. (\ref{eq:noisemeans1}, \ref{eq:noisemeans2}) to fall into the same form for both the Kondo lattice model and the model of the main text, just like the collision integrals in Eq. (\ref{eq:Cintuniversal}). While Eq. (\ref{eq:noisemeans3}) doesn't contain the integrals over $\hat{\mathbf{k}}'$ required to apply Eq. (\ref{eq:reduction}) and immediately reduce it to the same form as the main text, we note from the analysis in Sec. \ref{sec:gp} that the $\hat{\mathbf{k}}$ dependence of the distribution functions on the RHS of Eq. (\ref{eq:noisemeans3}) doesn't matter when $L\gg l$. Since the remaining non-$\hat{\mathbf{k}}$ dependent part of the distribution function is the same for $f_c$ and $f_f$, the contribution of Eq. (\ref{eq:noisemeans3}) to the noise power $S$, and hence the total $S$, is therefore equivalent to that in the main text. The results for the Fano factor are therefore the same for the Kondo lattice model and the simple $g'$ model of the main text.     

\section{Holstein phonons}
\label{app:HP}

\begin{figure}[h]
\begin{minipage}[h]{0.45\linewidth}
\center{\includegraphics[width=1\linewidth]{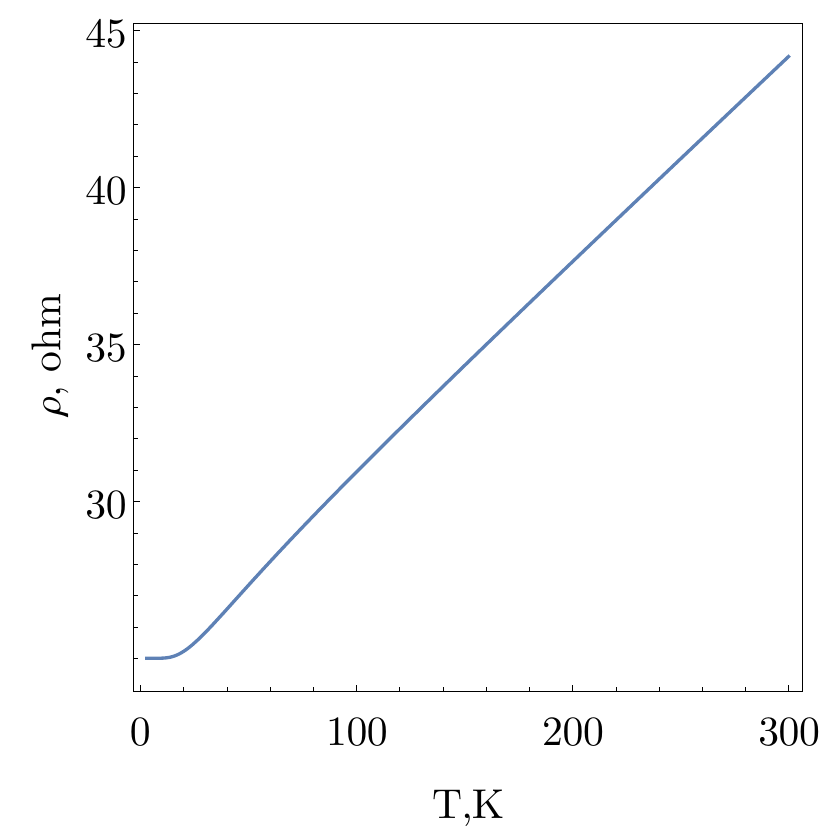}}
\\(a)
\end{minipage} 
\hfill    
\begin{minipage}[h]{0.45\linewidth}
\center{\includegraphics[width=1\linewidth]{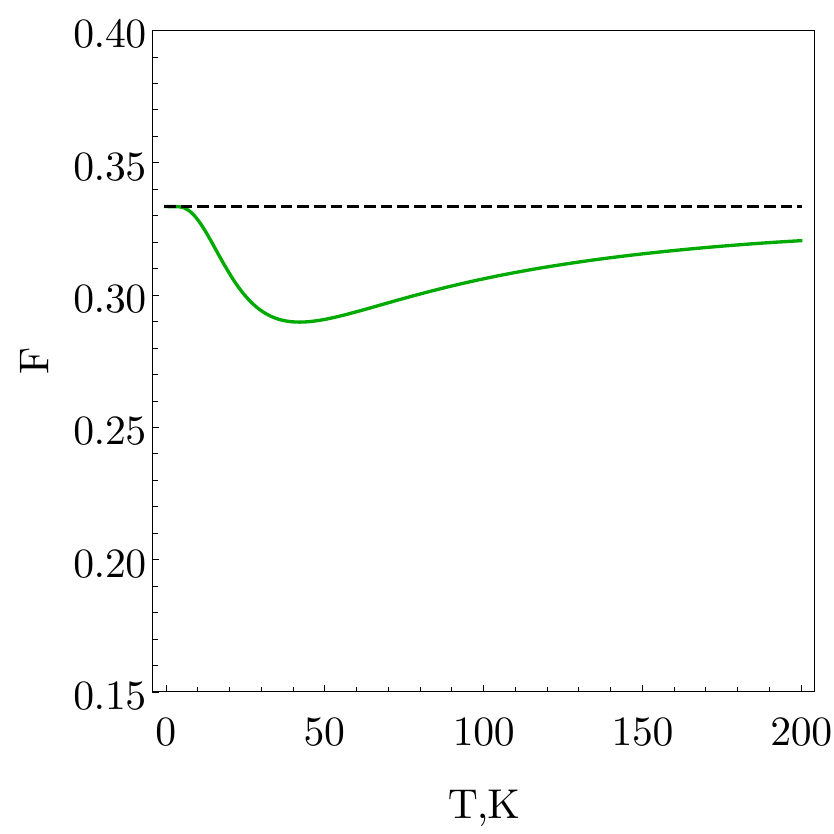}}
 \\(b)
\end{minipage} 
\caption{(a) Resistance as a function of temperature for Holstein phonon scattering. \
(b) Fano factor as a function of temperature, computed using Eq. (\ref{eq:FFS}). Parameters: $V=60~\mathrm{K}$, $\lambda/v=0.9$, $m_B=100~\mathrm{K}$. The remaining parameters are chosen to get a residual resistance of $25~\Omega$ like in the gold nanowire of Ref. \cite{Natelson23}.}
\label{fig:shot_noise_phonons}
\end{figure}

In this Appendix we compute the shot noise for the diffusive metal interacting with Holstein phonons and compare it with our main result. The retarded Green's function of the dispersionless phonons is $D_{\mathrm{loc}}^{R}(\Omega)=(1/(2m_B))(1/(\Omega+m_B+i0^+)-1/(\Omega-m_B+i0^+))$, which results in
\begin{equation}
B_{\mathrm{loc}}(\Omega)=\frac{\pi}{m_B}\left[\delta(\omega-m_B)-\delta(\omega+m_B)\right].
\end{equation}
Using Eq. (\ref{eqn:s_omega}) for $s(\omega)$ we obtain
\begin{equation}
 s(\omega)=-\frac{2\pi v_F}{(2 \pi)^2 m_B v^2 } \left(n_B(m_B)-n_B(-m_B)+n_F(\omega+m_B)-n_F(\omega-m_B) \right).
\end{equation}
Most of the steps of the original calculations remain unchanged and we can directly compute the conductivity and the shot noise using Eqs. (\ref{sigmagp}, \ref{eq:mainresult}) respectively. Fig. \ref{fig:shot_noise_phonons} (a) shows the resistance as a function of temperature; for $T \gg m_B$ it becomes linear in $T$. Using the data from Ref. \cite{Natelson23} for resistance of the gold nanowire that shows similar behavior, we can extract the parameters $m_B=100~\mathrm{K}$ and $\lambda/v=0.9$, where $\lambda=g'\sqrt{\pi/m_B}$ is the electron-phonon coupling.

In the case of $T=0$, it is possible to obtain an analytic answer for the shot noise power in three cases
\begin{equation}
S=\frac{\sigma_r W}{L}
 \begin{dcases}
&\frac{2 V}{3}, V<m_B\\
&\frac{2 }{3}\left( m_B+\frac{V-m_B}{(1+2\lambda^2 /(2\pi v)^2)^2}\right)+\frac{2\lambda^2 }{3(2\pi)^2 v^2}  \left( \frac{V-m_B}{(1+2\lambda^2 /(2\pi v)^2)^2} +V-m_B\right), m_B<V<2m_B\\
&\frac{2 }{3}\left( m_B+\frac{V-m_B}{(1+2\lambda^2 /(2\pi v)^2)^2}\right)+\frac{2\lambda^2}{3(2\pi)^2 v^2} \left( \frac{2V-3m_B}{(1+2\lambda^2 /(2\pi v)^2)^2} +m_B\right), V>2m_B\\
 \end{dcases}
\end{equation}

In the case of $V\ll m_B$, $F=1/3$, and when $V \gg m_B$ the Fano factor is
\begin{equation}
F = \frac{S}{2 \sigma_r V W/L}=\frac{1}{3} \frac{1}{1+2\lambda^2 /(2\pi v)^2}.
\end{equation}
We note that, contrary to the $v-g'$ model of the strange metal (Eq. (\ref{eqn:fano_factor})), the Fano factor does not reach a universal value and can be arbitrarily small when both the interaction $\lambda$ and the voltage $V$ are large enough. The experiment on the gold nanowire in Ref. \cite{Natelson23} was done for $V$ up to $70 ~\mathrm{K}$, so we are in the regime where $V<m_B$, and therefore $F(T=0)=1/3$, as is also observed.

Fig. \ref{fig:shot_noise_phonons} (b) demonstrates the Fano factor at finite temperature for the choice of parameters corresponding to the gold nanowire experiment of Ref. \cite{Natelson23}. At $T=0$ inelastic scattering is not possible due to the large boson gap, and therefore $F=1/3$. In the intermediate regime $T \sim m_B$ the Fano factor experiences a dip since inelastic scattering takes place in the transport. However, it goes back to $1/3$ at $T \gg m_B$, because the electron-phonon scattering becomes quasi-elastic in this limit, and it is therefore indistinguishable from electron-impurity scattering (which also produces $F=1/3$) within the purview of shot noise. Overall, our prediction for $F(T)$ approximately matches the data for the gold nanowire in Ref. \cite{Natelson23} over the temperature range reported by them ($T<20~\mathrm{K}$), and uses no additional free parameters.

\bibliography{refs.bib}

\end{document}